\documentclass[a4paper,11pt]{article}
\pdfoutput=1 
\usepackage{jheppub} 
\usepackage[T1]{fontenc} 
\usepackage{amsmath}
\usepackage{amssymb}
\usepackage{amsfonts}
\usepackage{mathtools}
\usepackage{microtype}
\usepackage{tikz}
\usepackage{bm}
\usepackage[utf8]{inputenc}
\usepackage{blkarray}
\usepackage{bigstrut}
\usepackage{nccmath}
\usepackage{enumitem}
\usepackage{braket}
\usepackage{dsfont}
\usepackage{lipsum}
\usepackage{array,multirow}
\usepackage{physics}
\usepackage{xcolor}
\usepackage{color}
\usepackage{float}
\usepackage{subfig}
\usepackage[export]{adjustbox}
\usepackage{graphicx}
\usepackage{caption}
\usepackage{comment}
\usepackage{lscape}

\allowdisplaybreaks

\usepackage{xargs}
\usepackage[colorinlistoftodos,prependcaption,textsize=tiny]{todonotes}
\newcommandx{\will}[2][1=]{\todo[linecolor=purple,backgroundcolor=purple!25,bordercolor=purple,#1]{\underline{WJT}: #2}}

\def\una{ \left( \frac{\alpha^{\rm b}_{s}}{\pi} \right) }
\def\FeynArts{{{\sc FeynArts}}}
\def\FeynCalc{{{\sc FeynCalc}}}

\def\Aida{{{\sc Aida}}}
\def\Reduze{{{\sc Reduze}}}
\def\Kira{{{\sc Kira}}}
\def\Ginac{{{\sc Ginac}}}
\def\ep{{{\epsilon}}}

\newcommand{\munich}{Max-Planck-Institut f\"ur Physik, Werner-Heisenberg-Institut,\\
F\"ohringer Ring 6, D-80805 M\"unchen, Germany. 
}

\newcommand{\unipd}{Dipartimento di Fisica e Astronomia, Universit\`a degli Studi di Padova,\\ 
Via Marzolo 8, I-35131 Padova, Italy.}

\newcommand{\pdinfn}{INFN, Sezione di Padova,\\
Via Marzolo 8, I-35131 Padova, Italy.}

\newcommand{\napoliall}{Dipartimento di Fisica, Universit\`a di Napoli Federico II, and INFN, Sezione di Napoli,\\
Via Cinthia, I-80126 Napoli, Italy.}

\setlist[itemize,1]{leftmargin=1.5em}

\title{\boldmath Two-loop scattering amplitude for heavy-quark pair production through light-quark annihilation in QCD }

\author[a]{Manoj K. Mandal,}
\author[a,b]{Pierpaolo Mastrolia,}
\author[c]{Jonathan Ronca,}
\author[d]{and William~J.~Torres~Bobadilla}

\affiliation[a]{\pdinfn}
\affiliation[b]{\unipd}
\affiliation[c]{\napoliall}
\affiliation[d]{\munich}

\emailAdd{manojkumar.mandal@pd.infn.it}
\emailAdd{pierpaolo.mastrolia@unipd.it}
\emailAdd{ronca@na.infn.it}
\emailAdd{torres@mpp.mpg.de}

\preprint{MPP-2022-38}

\abstract{We present the first full analytic evaluation of the scattering amplitude for the process $q {\bar q} \to Q {\bar Q}$ 
up-to two loops in Quantum Chromodynamics, for a massless $(q)$ and a massive $(Q)$ quark flavour.
The interference terms of the one- and two-loop amplitudes with the Born amplitude, decomposed in terms of gauge invariant form factors depending on the colour and flavour structure, are analytically calculated by keeping complete dependence on the squared center-of-mass energy, the squared momentum transfer, and the heavy-quark mass. The results are expressed as Laurent series around four space-time dimensions, with coefficients given in terms of generalised polylogarithms and transcendental constants up-to weight four. 
Our results validate the known, purely numerical calculations of the squared amplitude, and extend the analytic knowledge, previously limited to a subset of form factors, to their whole set, coming from both planar and non-planar diagrams, up-to the second order corrections in the strong coupling constant.
}

\begin{document} 

\maketitle

\section{Introduction}
\label{sec:introduction}

The production of top quark $(t)$ and its anti-particle $(\bar t)$ has been occupying a central role within the precision physics programme at hadron colliders, like the Tevatron and the Large Hadron Collider (LHC), 
over the last three decades. Being the heaviest known elementary particle, the $t$ quark has offered a portal to the discovery of the Higgs boson, and it is considered pivotal for understanding the 
electroweak symmetry breaking mechanism. Studies of top-quark production (and decay)  
at the current LHC physics programme enters the high-precision tests of the parameters of the Standard Model (SM), such as couplings and masses, as well as the analyses of backgrounds, for discriminating deviations that could indicate the path to move beyond {\it it}. 
Within SM, the production of $t {\bar t}$ pairs in hadronic collisions is the main source of top quarks, therefore, it is considered among the cornerstone processes at the current and future hadron colliders. Because of its role for the precision physics programme at hadron colliders, the
$t {\bar t}$-pair production has triggered a significant progress in the developments of theoretical methods for determination of the (differential) cross-sections, hence it has been stimulating the constant effort of providing calculations in Perturbative Quantum Chromodynamics (QCD), of increasing order in the strong-coupling series expansion.

The cross-section for $t {\bar t}$-production at LHC, at leading order (LO) and next-to-leading order (NLO) in QCD has been known since long~\cite{Nason:1987xz, Beenakker:1988bq, Beenakker:1990maa, Nason:1989zy,Czakon:2008ii}. 
The total cross section up-to the next-to-next-to-leading order (NNLO) in QCD 
became available in~\cite{Barnreuther:2012wtj, Czakon:2012zr, Czakon:2012pz, Czakon:2013goa}.
Fully differential NNLO calculations require a major control over infrared (IR) divergences appearing at intermediate stages of the calculation.
Partial results were obtained by using the antenna subtraction method \cite{GehrmannDeRidder:2005cm, Abelof:2011jv,Abelof:2014fza, Abelof:2014jna,Abelof:2015lna}.
The complete NNLO predictions were first carried out in~\cite{Barnreuther:2012wtj, Czakon:2012zr, Czakon:2012pz,Czakon:2013goa,Czakon:2014xsa,Czakon:2015owf,Czakon:2016ckf,Czakon:2017dip}, by using the Stripper approach  \cite{Czakon:2010td,Czakon:2011ve,Czakon:2014oma}. 
More recently, the NNLO computation of heavy-quark hadroproduction has been also completed in~\cite{Bonciani:2015sha,Catani:2019hip,Catani:2019iny,Catani:2020tko,Catani:2020kkl},
within the $q_T$-subtraction scheme~\cite{Catani:2007vq}. 
For recent studies on the strategies to perform precise higher-order computations in high-energy physics, 
see Refs.~\cite{TorresBobadilla:2020ekr,Heinrich:2020ybq}.

The calculation of the NNLO QCD corrections to $pp \to t {\bar t}$ requires four types of terms:
the {\it double-real} corrections, coming from the tree-level squared amplitude 
for a process with two additional partons in the final state;
the {\it real-virtual} corrections, due to 
the interference of the tree-level and of the 
one-loop amplitude for a process with one additional gluon in the final state; 
the {\it squared one-loop} corrections; 
the {\it double-virtual} corrections, 
due to the interference of the two-loop amplitude with the tree-level one.

The scattering amplitude for the real-virtual contributions were evaluated in~\cite{Dittmaier:2007wz,Dittmaier:2008uj},
and more recently in~\cite{Badger:2022mrb}.
The purely virtual contributions depend on 
the square of one-loop amplitude and
the genuine two-loop amplitude.
 The former has been computed analytically in~\cite{Korner:2008bn,Anastasiou:2008vd,Kniehl:2008fd}, 
 while the latter has been determined completely numerically
in~\cite{Czakon:2008zk,Baernreuther:2013caa,Chen:2017jvi}.
The analytic evaluation of the two-loop amplitude is known partially~\cite{Bonciani:2008az,Bonciani:2009nb,Bonciani:2010mn,vonManteuffel:2013uoa,Bonciani:2013ywa,Badger:2021owl}.
The main difficulty, in this case, is due to the analytic evaluation of the independent  integrals appearing in the decomposition of the two-loop amplitudes, known as {\it master integrals} (MIs).

At parton-level, the $t {\bar t}$-production proceeds {\it via} the  
annihilation of a light-quark ($q$) and an anti-quark (${\bar q}$), 
$q {\bar q} \to t {\bar t}$,
and the more luminous gluon-fusion channel, $g g \to t {\bar t}$. \\
As regarding the gluon-fusion channel, the analytic evaluation of the interference of the two-loop amplitude with the tree-level amplitude is only partially complete, and they are expressed in terms of generalised polylogarithms (GPLs) and elliptic integrals~\cite{Czakon:2007wk,Bonciani:2010mn,vonManteuffel:2013uoa,Bonciani:2013ywa,Adams:2018bsn,Adams:2018kez}.
Very recently, the two-loop helicity amplitudes 
for the $t\bar{t}$-production in the gluon-fusion channel within the {\it leading colour approximation}, including the contribution of closed loops of quarks, 
has been computed in~\cite{Badger:2021owl}. \\
As regarding the light-quark pair annihilation channel, 
the interference of the two-loop amplitude with the corresponding tree-level amplitude can be decomposed in terms of ten form factors, according to the colour and flavour structure.
Eight of them are known analytically, and expressed in terms of  GPLs~\cite{Czakon:2007ej,Bonciani:2008az,Bonciani:2009nb}.
\\ 

In this work, 
we present the complete analytic evaluation of the two-loop scattering amplitude for the scattering process $q {\bar q} \to Q {\bar Q}$, with a massless ($q$) and a massive ($Q$) quark flavour, in QCD, including leading and sub-leading colour contributions. 
We calculate the whole set of ten form factors analytically, including the two form factors previously unavailable, which
take contribution from both planar and non-planar graphs. The latter do not contribute to the eight form factors already known, and their evaluation constitute part of the novel insights of the current work.

The loop integrals appearing in the un-renormalised interference terms of the one- and two-loop bare amplitudes with the leading-order one 
are regulated within the Conventional Dimensional Regularisation (CDR), where $d$ is the number of continuous space-time dimensions. 

The calculation is automated within the 
\Aida{}~\cite{Mastrolia:2019aid} framework, 
implementing the adaptive integrand decomposition algorithm~\cite{Mastrolia:2016dhn,Mastrolia:2016czu} and
interfaced:
to  \FeynArts~\cite{Hahn:2000kx}, \FeynCalc~\cite{Shtabovenko:2016sxi}, for the automatic diagram generation and algebraic manipulations of the integrands;
to \Reduze~\cite{vonManteuffel:2012np},
and \Kira~\cite{Maierhoefer:2017hyi}, for the generation of the relations required for the decomposition in terms of MIs;
to {\sc SecDec}~\cite{Borowka:2015mxa}, for the numerical evaluation of MIs, if needed;
to {\sc PolylogTools}~\cite{Duhr:2019tlz}, 
{\sc Ginac}~\cite{Vollinga:2004sn}, and 
{\sc HandyG}~\cite{Naterop:2019xaf}, 
for the numerical evaluation of the analytic expressions.
The cancellation of the ultraviolet (UV) divergences of the bare interference terms at one and two loops 
is carried out by renormalising the quark fields and masses in the  
{\it on-shell} scheme, and the strong coupling in the $\overline{\text{MS}}$-scheme, along the lines of \cite{Czakon:2008zk,Bonciani:2008az}.
By using the analytic expressions of the MIs~\cite{Gehrmann:1999as,Bonciani:2003te,Bonciani:2003hc,Bonciani:2008az,Mastrolia:2017pfy,DiVita:2019lpl,Becchetti:2019tjy},
the renormalised interference terms are finally expressed as Laurent series around $d=4$ dimensions, by keeping the complete dependence on the Mandelstam invariants $s$ and $t$, and on the heavy-quark mass $M$. 
The one- and two-loop contributions 
are computed, respectively, up-to the first-order term, and up-to the finite term, in the four dimensional series expansion, whose coefficients are expressed in terms of GPLs and transcendental constants of up-to weight four. 
The analytic results are obtained in a non-physical region, where the variables $s$ and $t$ are negative, and are  numerically continued to the physical region, above the heavy-quark pair-production threshold, $s \ge 4M^2$.

The structure of the infrared (IR) singularities of the massless and massive gauge theory scattering amplitudes has been studied in~\cite{Catani:1998bh,Sterman:2002qn,Aybat:2006mz,Aybat:2006wq,Gardi:2009qi,Gardi:2009zv,Becher:2009cu,Becher:2019avh,Becher:2009qa,
Mitov:2006xs,
Becher:2009kw}.
In the current work, the IR singularities of the two-loop renormalised amplitude are successfully compared to the predicted expression built within the Soft Collinear Effective Theory (SCET), along the lines of the method presented in~\cite{Becher:2009qa, Becher:2009kw} 
and~\cite{Ferroglia:2009ep,Ferroglia:2009ii}.

The study of the virtual NNLO QCD corrections for the process 
$q \bar q \to Q {\bar Q}$, hereby presented, extends to the non-Abelian case the study of the four-fermion scattering amplitude with one massive fermion pair, 
in gauge theories, recently completed for the process $e^+~e^-~\to~\mu^+~\mu^-$ in Quantum Electrodynamics (QED)~\cite{Bonciani:2021okt}. 

In the following pages, we describe 
the strategy we adopted to solve the problem of  
{\it the} analytic evaluation of the double-virtual NNLO corrections to one, out of two, partonic reactions contributing the hadroproduction of heavy-quark pair. Thus, providing what we consider an important
validation and extension of the purely numerically known results, which have been employed 
 to obtain state-of-the-art perturbative predictions 
 within {\it top}-quark physics studies at hadron colliders
 (see~\cite{Mazzitelli:2021mmm,ATLAS:2022xfj} and reference therein, for recent applications).

\section{Scattering Amplitude}
\label{sec:amplitude}
We consider the scattering amplitude of the process, 
\begin{equation}
	q(p_1) + \bar{q}(p_2) \rightarrow Q(p_3) + \bar{Q}(p_4)\, ,
\label{eq:qqttbar}	
\end{equation}
where $q \ [{\bar q}]$ stands for a massless quark [anti-quark], {\it i.e.} $m_q=0$, and 
$Q \ [{\bar Q}]$, for a massive quark [anti-quark], {\it i.e.} $m_Q = M \ne 0$, in QCD.
The Mandelstam invariants of the scattering reaction are
$	s=(p_1+p_2)^2$, $\, \, t=(p_1-p_3)^2, $ and $ \, u=(p_2-p_3)^2 $, satisfying the condition $s + t + u = 2 M^2$. 
In the physical region, the range of Mandelstam variables reads,
\begin{equation}
    s \ge 4M^2  \quad \& \quad  
    -\bigg({\sqrt{s} - \sqrt{s-4 M^2}\over 2}\bigg)^2
    \le t \le 
    -\bigg({\sqrt{s} + \sqrt{s-4 M^2}\over 2}\bigg)^2\,.
\end{equation}
The dependence of the scattering amplitude on the kinematic variables can be conveniently parametrised in terms of the dimensionless variables, $\eta$ and $\phi$, defined as,
\begin{equation}
    \eta = \frac{s}{4M^2} - 1\, ,\qquad \phi = \frac{M^2 - t}{s}\, ,
\end{equation}
which, in the physical region 
satisfy the conditions,
\begin{align}
    \label{eq:physreg}
    \eta > 0 \, \quad \& \quad
    \frac{1}{2}\left(1-\sqrt{\frac{\eta}{1+\eta}} \, \right) \, \leq \, \phi \, \leq \, \frac{1}{2}\left(1+\sqrt{\frac{\eta}{1+\eta}} \, \right)\, .
\end{align}

The scattering amplitude ${\cal A}$ of the process can be evaluated in perturbative QCD, and expressed as a power series in the strong coupling $\alpha_s$, as, 
\begin{eqnarray}
   \label{eq:twoloopR}
	{\cal A} \left(\alpha_s \right)  
	&=&  
	4 \pi \alpha_s
	\bigg[
	{\cal A}^{(0)}
	+ \left(\frac{\alpha_s}{\pi}\right) {\cal A}^{(1)} 
	+ \left(\frac{\alpha_s}{\pi}\right)^2 {\cal A}^{(2)} 
	+ {\cal O}\left(\alpha_s^3\right)
	\bigg] \ .
\end{eqnarray}
%

\begin{figure}[t]
\centering
\includegraphics[scale=0.65]{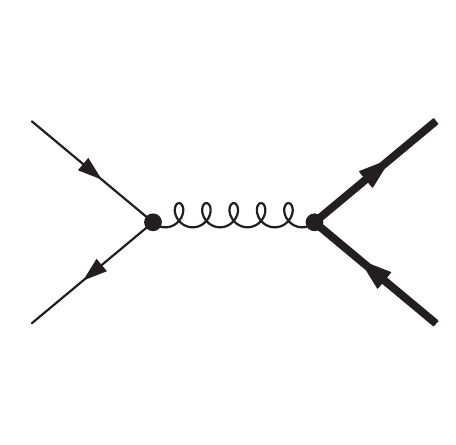}
\caption{Tree-level Feynman diagrams for the process $q\bar{q}\to Q\bar{Q}$. Thin lines indicate a light quark ($q$), whilst tick ones indicate a heavy quark ($Q$); curly lines correspond to gluons.}
\label{fig:treediag}
\end{figure}
\begin{figure}[t]
 \centering
 \includegraphics[scale=0.55]{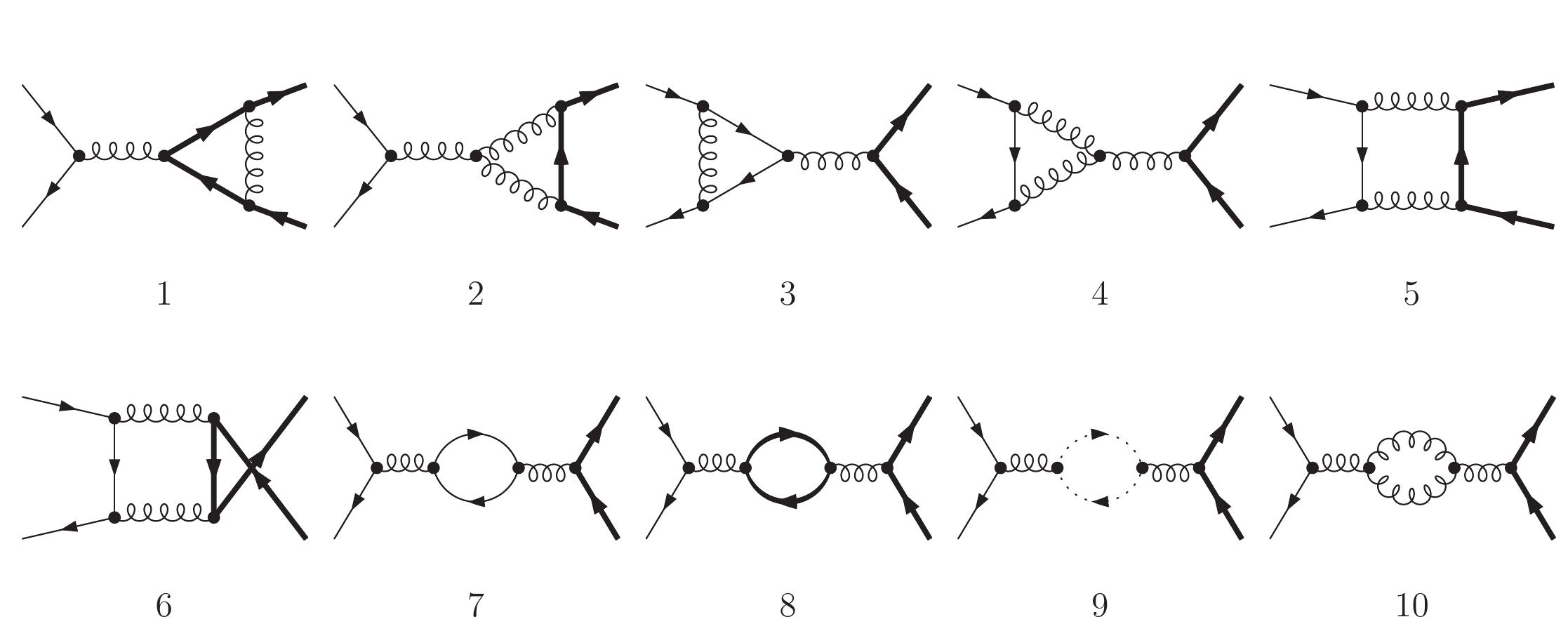}
 \caption{One-loop Feynman diagrams for the process $q\bar{q}\to Q\bar{Q}$. Thin lines indicate a light quark ($q$), whilst tick ones indicate a heavy quark ($Q$); curly and dashed lines correspond
 to gluons and ghosts, respectively.}
 \label{fig:oneloopdiagsall}
 \end{figure}
\begin{figure}[t]
 \centering
 \makebox[\textwidth][c]{\includegraphics[scale=0.38]{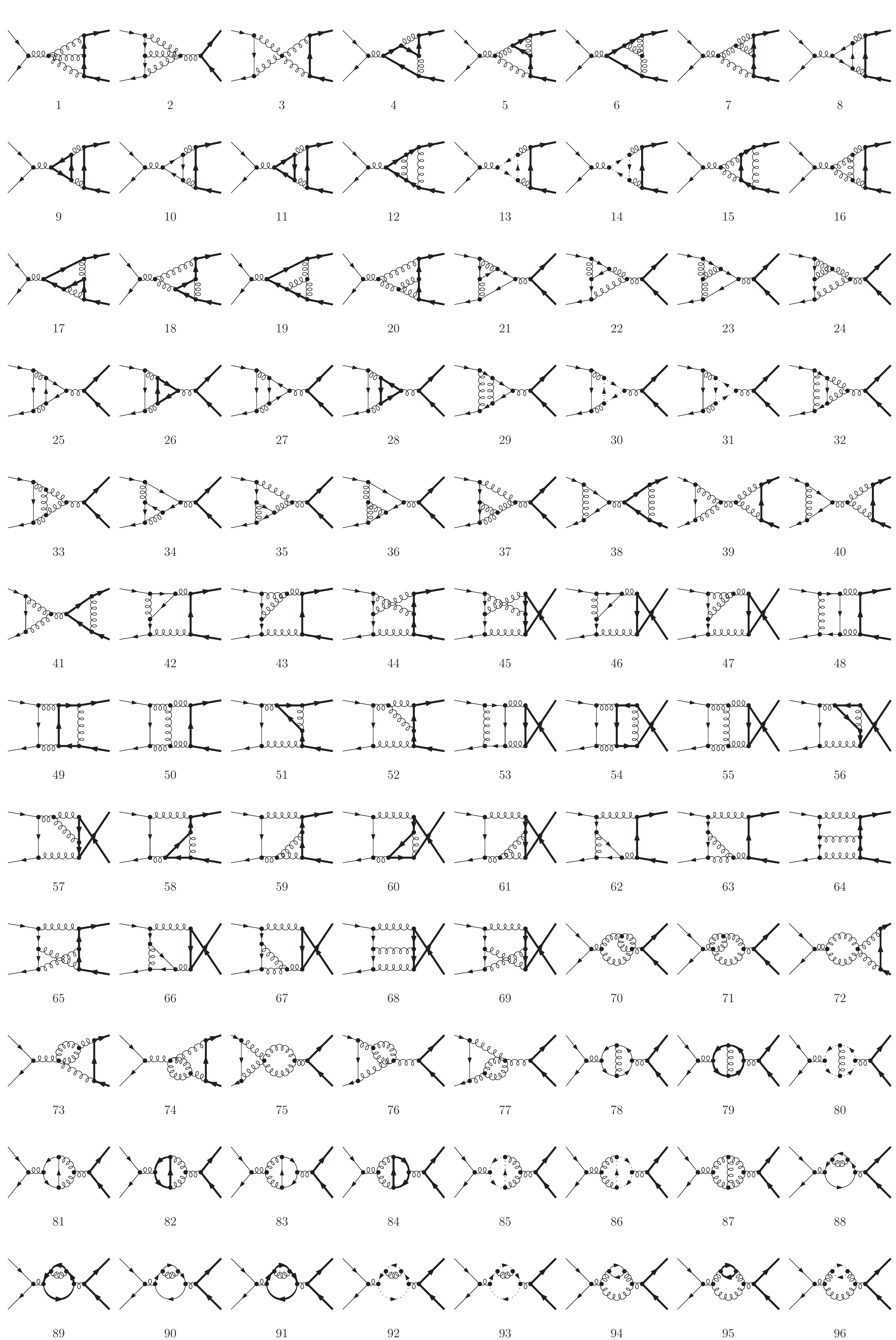}}
 \caption{Two-loop Feynman diagrams for the process $q \bar q \to Q \bar Q$ (set 1 of 2).}
 \label{fig:2Ldiagrams_part1}
\end{figure}
\begin{figure}[t]
 \centering
 \makebox[\textwidth][c]{\includegraphics[scale=0.38]{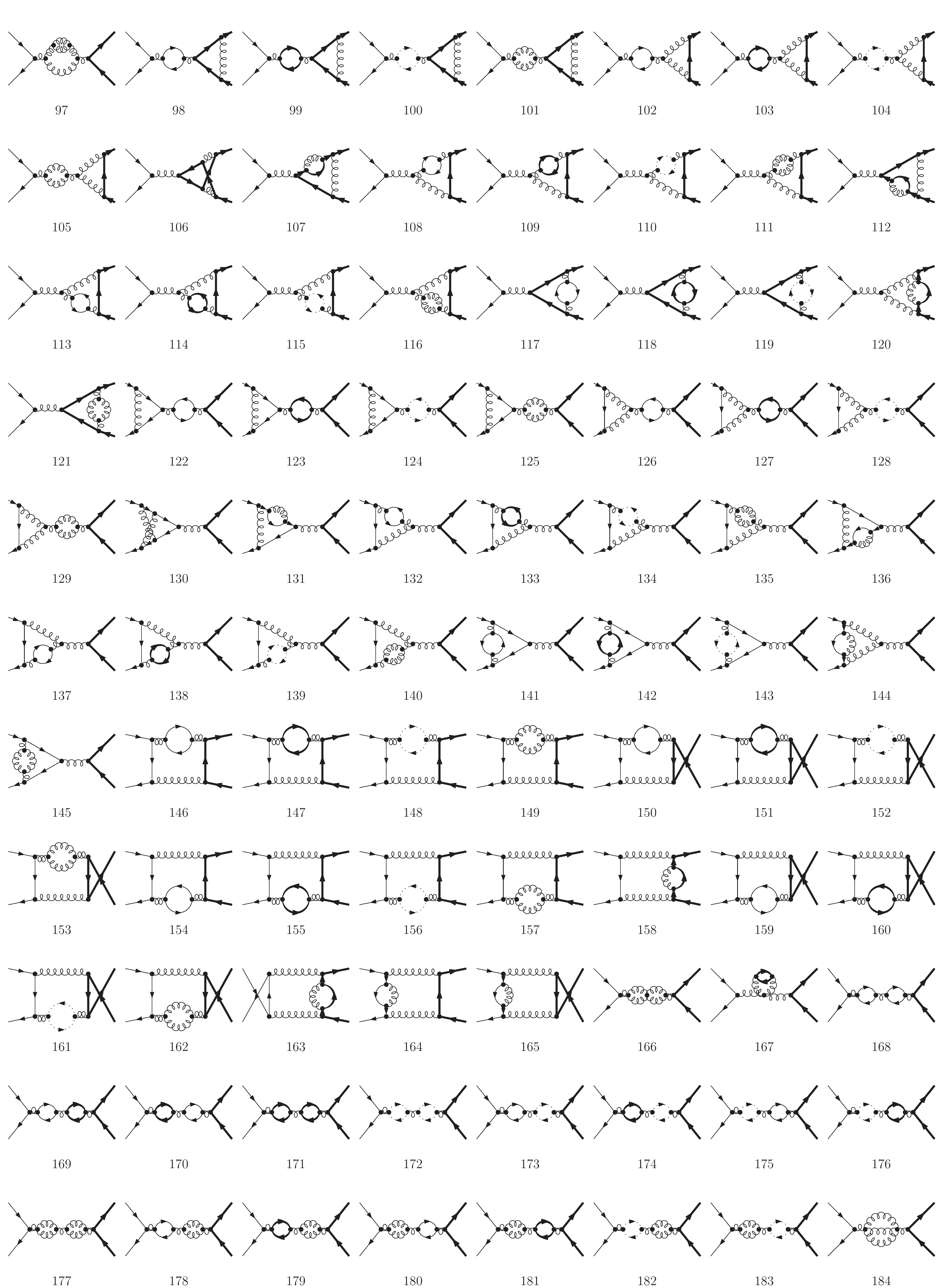}}
 \caption{Two-loop Feynman diagrams for the process $q \bar q \to Q \bar Q$ (set 2 of 2).}
 \label{fig:2Ldiagrams_part2}
 \end{figure}

The LO term 
${\cal A}^{(0)}$, 
referred to as {\it Born term},
receives contribution from a single tree-level Feynman diagram, see Fig.~\ref{fig:treediag}.
The colour-summed, un-polarised squared amplitude at LO 
(summed over the number of colours, summed over the final spins, and averaged over the initial states) 
has a rather simple expression,
\begin{align}
{\cal M}^{(0)} &=\frac{1}{4}\sum_{\substack{\rm colours\\ \rm spins}} 
|{\cal A}^{(0)}|^2    = \left(N_{c}^{2}-1\right)A^{(0)}
   \,,
\label{eq:renborn}   
\end{align}
with, 
\begin{align}
A^{(0)} &= \frac{2(1-\epsilon)s^{2}+4\left(t-M^{2}\right)^{2}+4st}{s^{2}}\,,
\end{align}
where $N_c$ is the number of colours, 
and $\epsilon = (4 - d)/2$, with $d$ being the number of (continuous) space-time dimensions.
The higher order contributions ${\cal A}^{(n)}$, with $n=1,2$, 
get contributions from one- and two-loop diagrams, respectively, shown in Figs.~\ref{fig:oneloopdiagsall}
and~\ref{fig:2Ldiagrams_part1},~\ref{fig:2Ldiagrams_part2}. 
The interferences of one- and two-loop amplitudes with the 
Born term are defined as, 
\begin{equation}
{\cal M}^{(n)} = 
\frac{1}{4}\sum_{\substack{\rm colours\\ \rm spins}}  2\, \text{Re} ( {\cal 
A}^{(0)*} \, 
{\cal A}^{(n)} ) \,, \ \text{for $n=1,2$}
\, ,
\label{eq:reninterference}
\end{equation} 
and
can be organised as combinations 
of gauge invariant factors, according to the dependence on 
the number of colours ($N_c$) and on the flavour structure
(i.e., the number of light-  
and heavy-fermion closed loops, respectively,  $n_l$ and $n_h$).
In particular, the contributions at one- and two-loop admit the following decomposition~\cite{Czakon:2007ej,Czakon:2008zk,Baernreuther:2013caa},
\begin{align}
{\cal M}^{(1)}&=2\left(N_{c}^{2}-1\right)\Bigg(A^{\left(1\right)}\,N_{c}+\frac{B^{\left(1\right)}}{N_{c}}+C_{l}^{\left(1\right)}\,n_{l}+C_{h}^{\left(1\right)}\,n_{h}\Bigg)\,,
\label{eq:deco1L}
\\
{\cal M}^{(2)}&=2\left(N_{c}^{2}-1\right)\Bigg(A^{\left(2\right)}\,N_{c}^{2}+B^{\left(2\right)}+\frac{C^{\left(2\right)}}{N_{c}^{2}}+D_{l}^{\left(2\right)}\,N_{c}\,n_{l}+D_{h}^{\left(2\right)}\,N_{c}\,n_{h}\notag
\\
&\qquad\qquad\qquad+E_{l}^{\left(2\right)}\frac{n_{l}}{N_{c}}+E_{h}^{\left(2\right)}\frac{n_{h}}{N_{c}}+
F_{l}^{\left(2\right)}\,n_{l}^{2}+
F_{lh}^{\left(2\right)}\,n_{l}\,n_{h}+
F_{h}^{\left(2\right)}\,n_{h}^{2}\Bigg)\,.
\label{eq:deco2L}
\end{align}

The analytic expressions of the one-loop form factors have been known since long time~\cite{Nason:1987xz, Nason:1989zy, Beenakker:1988bq, Beenakker:1990maa,
Mangano:1991jk, Korner:2002hy, Bernreuther:2004jv,Czakon:2008ii}.

Regarding the two-loop form factors in the colour decomposition~\eqref{eq:deco2L}, 
contributions from the {\it leading colour} ($A_{}^{(2)}$), 
one closed fermionic loop ($D_l^{(2)}$, $D_h^{(2)}$, $E_l^{(2)}$, $E_h^{(2)}$),
and two closed fermionic loops ($F_l^{(2)}, F_{lh}^{(2)}, F_{h}^{(2)}$) 
are known both numerically as well as analytically~\cite{Czakon:2008zk,Bonciani:2008az,Bonciani:2009nb}; 
$B^{(2)}$ and $C^{(2)}$, instead, are known only numerically~\cite{Czakon:2008zk}. 
Their analytic evaluation requires the evaluation of non-planar diagrams 
(that give no contribution to the leading colour term), and they are presented for the first time in this work.

The evaluation of the previously known colour factors, together with the novel calculation of $B^{(2)}$ and $C^{(2)}$,
allows us to obtain, for the first time, the complete analytic expression of the two-loop scattering amplitude 
for the four-quark scattering in QCD with a massive quark-pair, 
both as internal and as external states.

The results for the four-quark scattering $q {\bar q} \to Q {\bar Q}$ in QCD, hereby presented, 
can be considered as the natural extension to a non-Abelian theory 
of the ones obtained for the four-fermion scattering $e^{+} e^{-} \to \mu^{+} \mu^{-}$ in QED, 
recently presented in~\cite{Bonciani:2021okt}.
We observe that the coefficient $C^{(2)}$, 
as well as $E_l^{(2)}, E_h^{(2)}$, $F_l^{(2)}, F_{lh}^{(2)}$, 
and $F_{h}^{(2)}$,
can be written as linear combination of (colour stripped) 
Feynman diagrams that appear also in the Abelian case.
The form factors $A^{(2)}$, $B^{(2)}$, $D_l^{(2)}$, and $D_h^{(2)}$ 
get contribution from Abelian and non-Abelian (colour stripped) diagrams.
We refer the Reader to Appendix~\ref{app:colordeco} for a detailed discussion
on the colour decomposition. 

The complete analytic calculation of ${\cal M}^{(2)}$, 
or in other words, the computation of the form factors in decomposition~\eqref{eq:deco2L},
is the main result of the present manuscript. 

\section{Analytic Evaluation}
\label{sec:analyticevaluation}
We begin by considering 
the bare LO squared amplitude and the bare interference terms, 
\begin{eqnarray}
{\cal M}^{(0)}_{\rm b} &=&
\frac{1}{4}\sum_{\substack{\rm colours\\ \rm spins}}
|{\cal A}^{(0)}_{\rm b}|^2 \ , 
\label{eq:bareborn}   
\\
{\cal M}^{(n)}_{\rm b} &=& 
\frac{1}{4}\sum_{\substack{\rm colours\\ \rm spins}}
2\, \text{Re} ( {\cal \, A}^{(0)*}_{\rm b} \, 
{\cal A}^{(n)}_{\rm b} ) \,, \ \text{for $n=1,2$}
\, ,
\label{eq:barinterference}
\end{eqnarray} 
where ${\cal \, A}^{(n)}_{\rm b}$ $(n \ge 0)$ are 
the coefficients of the series expansion
of the bare amplitude ${\cal A}_{\rm b}$ in the bare strong coupling constant,
$\alpha^{\rm b}_{s} \equiv g_{s}^2/(4 \pi)$. Its expression up-to the second-order corrections
reads as,
\begin{align}
	{\cal A}_{\rm b}\left(\alpha^{\rm b}_{s}\right)
	=  
	4 \pi &\alpha^{\rm b}_{s}\, S_{\epsilon} \, \mu^{-2 \epsilon}
	\bigg[ 
	{\cal A}_{\rm b}^{(0)}
	+ \una {\cal A}_{\rm b}^{(1)} + \una^2 \!\! {\cal A}_{\rm b}^{(2)} 
	+ O\Big( (\alpha^{\rm b}_{s})^3\Big)
	\bigg] \,, 
   \label{eq:unrenormamp}	
\end{align} 
with    
$S_{\epsilon} \equiv (4\pi e^{-\gamma_{E}})^{\epsilon}$,
and $\mu$ being the 't Hooft mass scale.
The CDR scheme is adopted throughout the whole computation, 
hence, internal and external states are, accordingly, regularised
in $d=4-2\epsilon$ space-time dimensions~\cite{tHooft:1972tcz,Bollini:1972ui,Gnendiger:2017pys}.
The LO term ${\cal A}_{\rm b}^{(0)} = {\cal A}^{(0)}$,  given in Eq.~(\ref{eq:renborn}),
is finite in the limit $d\to4$ 
($\epsilon \to 0$);
whereas the higher order terms require the evaluation of one- and two-loop integrals that may contain UV and IR divergences, parametrised as poles in $\epsilon$.

\subsection{UV Renormalisation}
\label{sec:UVRenormalization}
The one- and two-loop amplitudes contain both UV and IR divergences. The UV divergences are removed by renormalising the bare quark fields and the bare mass of the heavy quark in the on-shell scheme, 
\begin{eqnarray}\label{qcdren}
	{\psi}_{\rm b} &=& \sqrt{Z_2} \, {\psi}\,, \quad
	 M_{\rm b} = Z_M M \, ,
\end{eqnarray}
and by renormalising the bare coupling constant $\alpha_s^{{\rm b}}$ at the scale $\mu$ in the $\overline{\text{MS}}$ scheme,
\begin{equation}
  \alpha_s^{{\rm b}}\,S_{\epsilon}=\alpha_s(\mu^{2})\,\mu^{2\epsilon}\,Z_{\alpha_s}^{\scriptscriptstyle\overline{\text{MS}}}\,.
  \label{eq:alphasrenorm}
\end{equation}
By employing this, we can express the renormalised amplitude in terms of the bare amplitude as,
\begin{eqnarray}
	\label{eq:renamp}
	{\cal A} 
	&=& Z_{2,q} \, Z_{2,Q} \, 
	{\cal A}_{\rm b} \Big(
\alpha_s^{\rm b} = \alpha_s^{\rm b}(\alpha_s)\,, 
M_{\rm b} = M_{\rm b}(M)
\Big)\ ,
\end{eqnarray}
where $Z_{2,q}$ and $Z_{2,Q}$ are the on-shell wave function renormalisation constants for the massless and massive quarks;  $M$ is the renormalised mass for the heavy quark in the on-shell scheme. The renormalised amplitude depends on four renormalisation constants ($Z_{2,q}, Z_{2,Q}, Z_{\alpha_s}, Z_M$), which admit a perturbative expansion in the renormalised coupling constant $\alpha_s$, 
\begin{align}
\label{eq:Zexpansion}
Z_{j} &= 1 + \left(\frac{\alpha_s}{\pi}\right) \delta Z_{j}^{(1)} + \left(\frac{\alpha_s}{\pi}\right)^2 \delta Z_{j}^{(2)} 
+ {\cal O}(\alpha_s^3) \ , \,\,\, \text{for} \,\,\, j=\{q,Q,\alpha_s, M\}\, .
\end{align}
The mass and wave-function renormalisation of the heavy quark is known to three loop accuracy in the on-shell scheme~\cite{Chetyrkin:1999ys,Melnikov:2000qh,Melnikov:2000zc}; the wave-function renormalisation of the light quark, due to the presence of heavy quark, is provided at two loop accuracy in~\cite{Czakon:2007ej}; the strong coupling constant renormalisation is known up-to five-loop accuracy~\cite{vanRitbergen:1997va,Czakon:2004bu,Baikov:2016tgj,Luthe:2016ima,Herzog:2017ohr,Chetyrkin:2017bjc}.
Their expressions, up-to the required order, are collected in Appendix~\ref{app:renconstants}.

\begin{figure}[t!]
\centering
\includegraphics[scale=0.45]{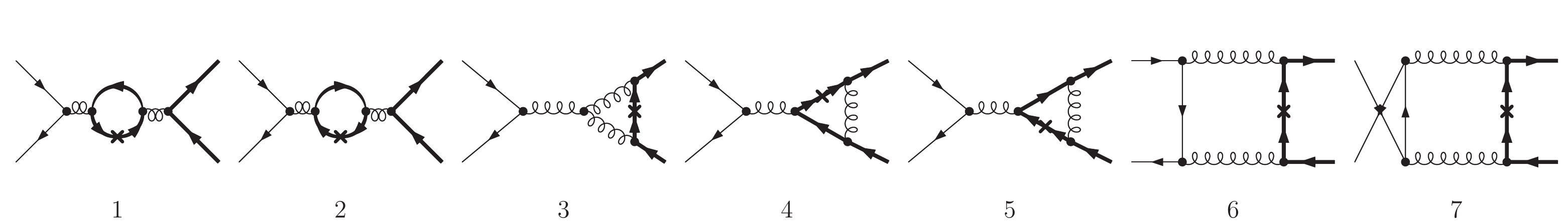}
\caption{Mass renormalisation counter-term diagrams.}
\label{fig:CTmass}
\end{figure}

Upon combining Eqs.~(\ref{eq:unrenormamp}),~(\ref{eq:renamp}), and~(\ref{eq:Zexpansion}), 
we obtain the UV renormalised amplitude ${\cal A}$, given in Eq.~(\ref{eq:twoloopR}),
%
%
whose coefficients ${\cal A}^{(n)}$ can be written in terms of the bare coefficients  
${\cal A}_{\rm b}^{(n)}$, as,
\begin{eqnarray}
	{\cal A}^{(0)} &=& {\cal A}_{\rm b}^{(0)}, \qquad 
	{\cal A}^{(n)}  = {\cal A}_{\rm b}^{(n)} + \delta {\cal A}^{(n)} \ , \quad (n>0)\,,
\label{eq:RenAmpInBareAmp}
\end{eqnarray}
with, 
\begin{subequations}
\begin{eqnarray}
	\delta {\cal A}^{(1)}  &=& 
	\Big( \delta Z_{\alpha_s}^{(1)}
	+ \delta Z_{Q}^{(1)} \Big) {\cal A}_{\rm b}^{(0)},
	\\
	\delta  {\cal A}^{(2)} &=& 
	\Big(2 \delta Z_{\alpha_s}^{(1)} + \delta Z_{Q}^{(1)} \Big) {\cal A}_{\rm b}^{(1)} 
	+\Big(\delta Z_{\alpha_s}^{(2)} 
	+ \delta Z_{Q}^{(2)} 
	+ \delta Z_{q}^{(2)} 
	+ \delta Z_{Q}^{(1)} \delta Z_{\alpha_s}^{(1)}
	\Big) {\cal A}_{\rm b}^{(0)} \nonumber \\
	&+&
	\delta Z_M^{(1)} 
	{\cal A}_{\rm b}^{(1, \text{mass CT})} \label{eq:aux1}
\, . 
\end{eqnarray}
\end{subequations}
The last term in Eq.~(\ref{eq:aux1}), corresponding to the mass renormalisation counter-term,
takes contributions from the diagrams depicted in Fig.~\ref{fig:CTmass} and
consists of the one-loop diagrams with an insertion of the mass counter-term in the heavy-quark propagators.

With the above definitions, one- and two-loop renormalised interference terms 
${\cal M}^{(n)}$ are obtained as,
\begin{eqnarray}
{\cal M}^{(n)} = {\cal M}^{(n)}_{\rm b} + \delta {\cal M}^{(n)}
\ , \qquad \ \text{for $n=1,2$} \ ,
\end{eqnarray}
where,
\begin{eqnarray}
\delta {\cal M}^{(n)} =
\frac{1}{4}\sum_{\substack{\rm colours\\ \rm spins}}
2\, \text{Re} 
\Big( {\cal \, A}^{(0)*}_{\rm b} \, 
\delta {\cal A}^{(n)} \Big) \ . 
\end{eqnarray}

\subsection{Algebraic decomposition}
\label{sec:analyticdeco}

\begin{figure}[h]
 \centering
 \includegraphics[scale=0.4]{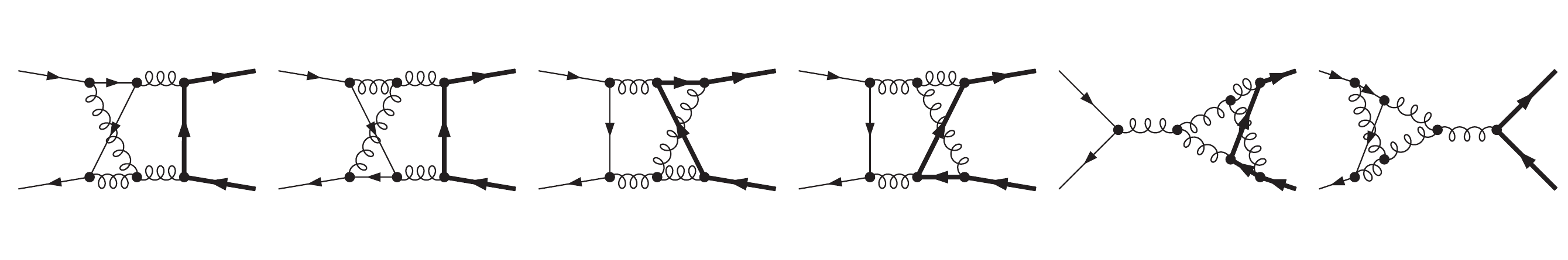}
 \caption{Two-loop diagrams that, upon interference with the Born amplitude,
 give rise to vanishing contributions,
 due to colour algebra.
 }
 \label{fig:zeronp}
 \end{figure}

The generation of the one- and two-loop diagrams contributing to 
${\cal M}^{(1)}_{\rm b}$ and ${\cal M}^{(2)}_{\rm b}$, as well as of those needed for the mass-renormalisation, 
is carried out using \FeynArts~\cite{Hahn:2000kx}. By choosing Feynman gauge for the gluon propagator,
we identify 10 diagrams at one loop, 184 diagrams at two loops, and 7 counter-term diagrams for the mass renormalisation,
respectively, shown in Fig.~\ref{fig:oneloopdiagsall}, 
Figs.~\ref{fig:2Ldiagrams_part1} and~\ref{fig:2Ldiagrams_part2},
and  Fig.~\ref{fig:CTmass}.
Scaleless loop integrals (e.g., one- and two-loop massless tadpoles),
and non-planar diagrams that vanish because of colour algebra 
(see Fig.~\ref{fig:zeronp}) are neglected.\footnote{Details on the diagrammatic contributions to the colour decomposition can be found in  Appendix~\ref{app:colordeco}.}

After performing colour, spin and Dirac-$\gamma$ algebra by means of \FeynCalc{}~\cite{Shtabovenko:2016sxi}, 
the interference terms are expressed in terms of $n$-loop scalar integrals as,
\begin{equation}
{\cal M}^{(n)}_{\rm b} = 
( S_{\epsilon} )^n
 \int \prod_{i=1}^n \frac{d^d k_i}{(2 \pi)^d} \, 
\sum_{G} \frac{N_G (p_i,k_i,M^2)}{\prod_{\sigma \in G} D_\sigma (p_i,k_i,M^2)} \quad , 
\end{equation}
where
$G$ denotes an $n$-loop graph interfered with the Born terms, 
$D_\sigma$ denotes the set of denominators corresponding to the internal lines of $G$,
and $N_G$ stands for a polynomial in the scalar products built out 
of external momenta $p_i$ and loop momenta $k_i$, and $M^2$.

\begin{figure}[h!]
\centering
\includegraphics[scale=0.4]{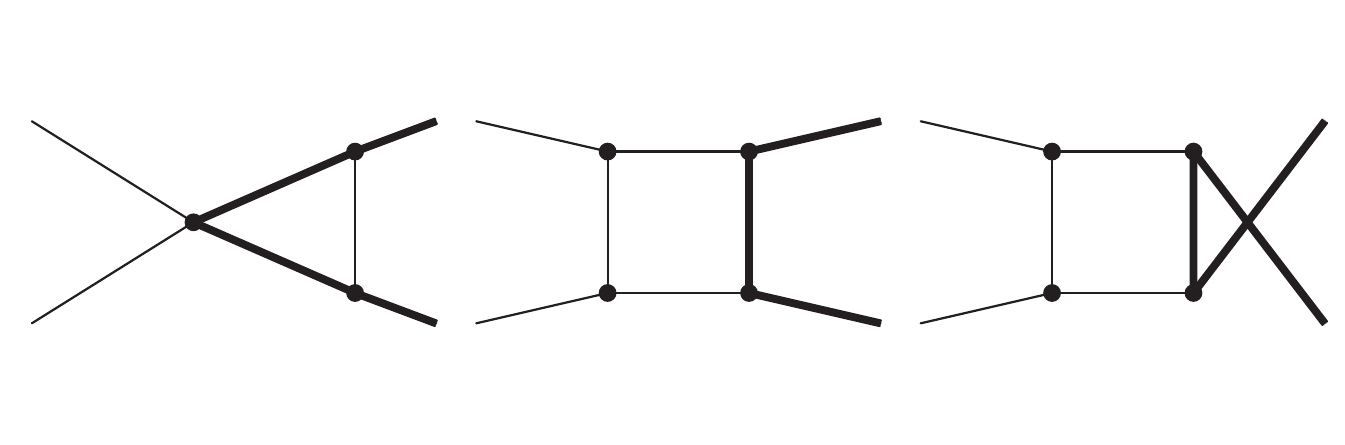}
\caption{One-loop parent graphs for the process $q\bar{q}\to Q\bar{Q}$. 
Thin lines indicate massless propagators, whilst tick ones indicate massive ones. 
}
\label{fig:parent_1L}
\end{figure}

\begin{figure}[t]
\centering
\includegraphics[scale=0.4]{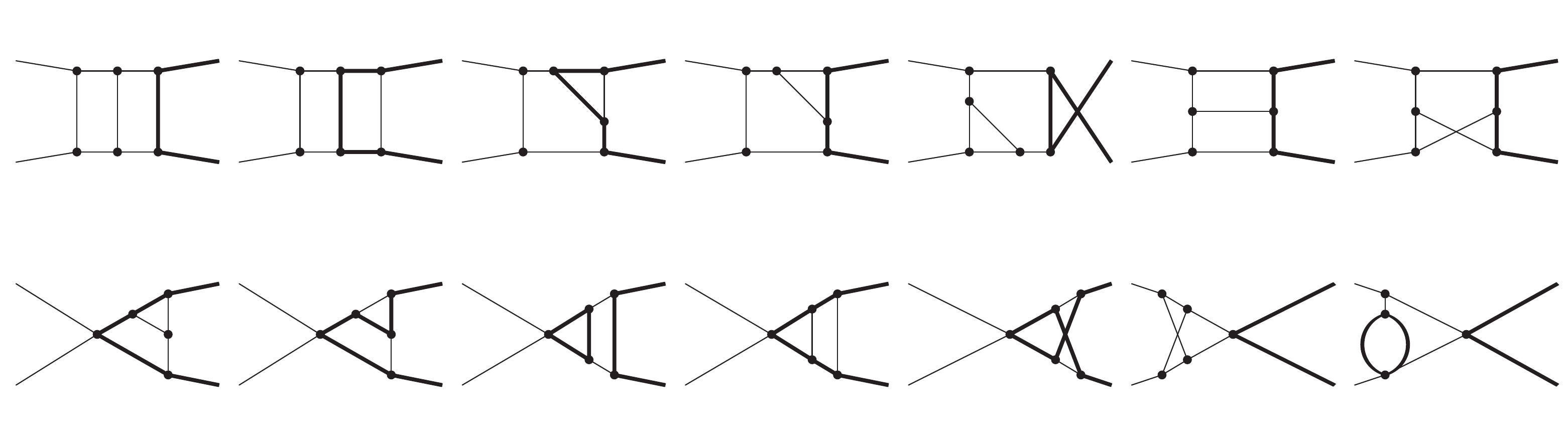}
\caption{Representative two-loop parent graphs for the process $q\bar{q}\to Q\bar{Q}$. 
Thin [Thick] lines indicate massless [massive] particles.}
\label{fig:parent_2L}
\end{figure}

The decomposition of the integrals is automated within the \Aida{} framework~\cite{Mastrolia:2019aid}, where integrands are grouped according to their common set of propagators with respect to the ones of the parent graphs, identified among all the diagrams as the ones with the largest sets of independent denominators.
At one-loop, {\sc Aida} identifies 3 parent graphs, shown in Fig.~\ref{fig:parent_1L};
at two-loop, 31 parent diagrams 
(22 belonging to four-point topologies and 9 belonging to three-point topologies), 
shown in Fig.~\ref{fig:parent_2L}, for representative topologies. 

The quantities ${\cal M}^{(n)}$ are simplified within \Aida{} by employing the {\it adaptive integrand decomposition method} \cite{Mastrolia:2016dhn,Mastrolia:2016czu} followed by the use of {\it integration-by-parts identities}~\cite{Tkachov:1981wb,Chetyrkin:1981qh,Laporta:2001dd}. The latter are automatically generated for the parent diagrams only, generated by {\sc Aida} through its interface to the public codes {\sc Reduze}~\cite{vonManteuffel:2012np} 
and {\sc Kira}~\cite{Maierhoefer:2017hyi}.
After integrand and integral decompositions,
the interference terms ${\mathcal{M}}_{\rm b}^{\left(n\right)}$
appear to be written as linear combinations of a set of independent MIs, 
say ${\bf I}^{(n)}$,
\begin{eqnarray}
    {\cal M}^{(n)}_{\rm b}  &=&  
    {\mathbb C}^{(n)} \, \cdot \,
    {\bf I}^{(n)} \, ,
\end{eqnarray}
where ${\mathbb C}^{(n)}$ represents a vector of coefficients, rational functions depending on $\epsilon$ and the kinematic variables, $s,t,M^2$. 
In particular, at one-loop,
${\bf I}^{(1)}$ is a vector of 12 MIs, 
and, at two-loop, ${\bf I}^{(2)}$ is a vector of 270 MIs, analytically known: 
two- and three-point functions, and a subset of the planar four-point functions have been known since long~\cite{Gehrmann:1999as,Bonciani:2003te,Bonciani:2003hc,Bonciani:2008az,Bonciani:2009nb},
whereas the complete set of planar and non-planar four-point integrals were computed 
in~\cite{Mastrolia:2017pfy,DiVita:2018nnh,DiVita:2019lpl}\footnote{
A comparison on a planar subset of master integrals, computed both in~\cite{Bonciani:2008az} and in~\cite{Mastrolia:2017pfy},
partly performed along the lines of~\cite{Henn:2021aco},
revealed that the numerical coefficient (a rational number) of $\pi^4$, 
within the weight-four term of the integrals $I_{30}$ and $I_{31}$, 
defined in Eq.~(6.2) of~\cite{Mastrolia:2017pfy}, 
was not correct. The revised version of the corresponding ancillary file, containing the analytic expression of the planar set of MIs used in this work, is available on the arXiv.} 
using the differential equation method via Magnus exponential~\cite{Argeri:2014qva}, 
and independently in~\cite{Becchetti:2019tjy}.

The one-loop counter-term 
$\delta {\cal M}^{(1)}$ is directly computed from the knowledge of the 
renormalisation constants $\delta Z_j$
and the Born squared amplitude.
Differently, the two-loop counter-term 
$\delta {\cal M}^{(2)}$ 
requires also the decomposition of one-loop integrals, due to both the genuine one-loop amplitude 
and to the mass renormalisation counter-term, 
coming from the one-loop diagrams shown in Fig.~\ref{fig:CTmass}, 
and, therefore, it admits a decomposition in terms of the basis ${\bf I}^{(1)}$.

\section{Results}
\label{sec:results}

After inserting the expression of the MIs and adding the bare quantities
${\cal M}^{(n)}_{\rm b}$ to the corresponding counter-terms
$\delta{\cal M}^{(n)}$, finally,
the renormalised interference terms 
${\cal M}^{(n)}$ are analytically  expressed as a Laurent series around $\epsilon =0$, as 
\begin{eqnarray}
 {\cal M}^{(1)} = \sum_{k=-2}^1 
 {\cal M}^{(1)}_{k} \, \epsilon^k \, 
 + {\cal O}(\epsilon^2)
 \, , \qquad {\rm and} \quad  
 {\cal M}^{(2)} = \sum_{k=-4}^0 
 {\cal M}^{(2)}_{k} \, \epsilon^k \, 
 + {\cal O}(\epsilon) \ ,
 \end{eqnarray}
whose coefficients ${\cal M}^{(n)}_{k}$ contain GPLs, iteratively defined as~\cite{Goncharov:1998kja},
\begin{subequations}
\begin{eqnarray}
  G(w_n,\ldots,w_1 ; \tau) 
  &\equiv&
  \int_0^\tau \frac{dt}{t-w_n} 
  G(w_{n-1},\ldots,w_1 ;t) \, , \\
  {\rm with} \quad 
  G(w_{1};\tau) &\equiv& \log(1-{\tau \over w_1}) \, . \quad
\end{eqnarray}
\end{subequations}

The analytical expression of ${\cal M}^{(1)}$ and ${\cal M}^{(2)}$ are computed in the non-physical region, $s < 0$, $t < 0$, and their analytic continuation to the region of heavy-quark pair production is performed numerically.
In particular, ${\cal M}^{(2)}$  contains 
5033 GPLs up-to weight four, whose arguments are written in terms of 18 letters, $w_i=w_i(x,y,z)$, which depend on the Mandelstam variables through the relations~\cite{DiVita:2018nnh,Mastrolia:2017pfy,DiVita:2019lpl},
\begin{equation}
   -\frac{s}{M^2} = x, \quad  -\frac{t}{M^2} = \frac{(1-y)^2}{y}, \quad  
   -\frac{u-M^2}{s-M^2}= \frac{z^2}{y} \ .
\end{equation}
%
The numerical evaluation of GPLs, in the physical region~\eqref{eq:physreg}, is performed by adopting the prescription,
\begin{equation}
    s \to s + i\delta 
    \ , 
    \label{eq:deltaim}
\end{equation} 
by assigning a small positive imaginary part to the squared center-of-mass energy variable, above the pair production threshold.\footnote{The numerical effect of $\delta \neq 0$ has been estimated to be of ${\cal O}(\delta)$, therefore, yielding numerical values of the interference terms in double precision with a choice of $\delta \sim {\cal O}(10^{-17})$.}


As anticipated in Sec.~\ref{sec:amplitude}, the analytic evaluation of the one-loop amplitude has been performed long ago
by following different approaches~\cite{Nason:1987xz,Nason:1989zy,Beenakker:1988bq,Beenakker:1990maa,Mangano:1991jk,Korner:2002hy,Bernreuther:2004jv,Czakon:2008ii}. 
On the two-loop side, instead, analytic expressions for the form factors 
present in the colour decomposition~\eqref{eq:deco2L}
is partially known.
In particular, the knowledge of these analytic expressions is restricted to leading-colour and closed fermion-loop terms 
($A_{}^{(2)}$, $D_l^{(2)}$, $D_h^{(2)}$, $E_l^{(2)}$, $E_h^{(2)}$, $F_l^{(2)}, F_{lh}^{(2)}, F_{h}^{(2)}$)~\cite{Bonciani:2008az,Bonciani:2009nb}.
The analytic evaluation of $B^{(2)}$ and $C^{(2)}$ required the evaluation of non-planar diagrams, 
that were absent from the leading-colour term, and they are presented for the first time in this work.

The independent evaluation of the previously known form factors, together with the novel calculation of $B^{(2)}$ and $C^{(2)}$, allows us to validate the previously known numerical results~\cite{Czakon:2008zk}, and to obtain, for the first time, the complete analytic expression 
of the two-loop scattering amplitude for the partonic scattering 
$q {\bar q} \to Q {\bar Q}$ in QCD.
Our result is the first example of a complete analytic calculation of a two-loop amplitude in QCD
with a massive quark-pair in the internal and as well as external states, including both the leading and sub-leading colour contributions. 

\begin{figure}
    \centering
    \includegraphics[scale=0.8]{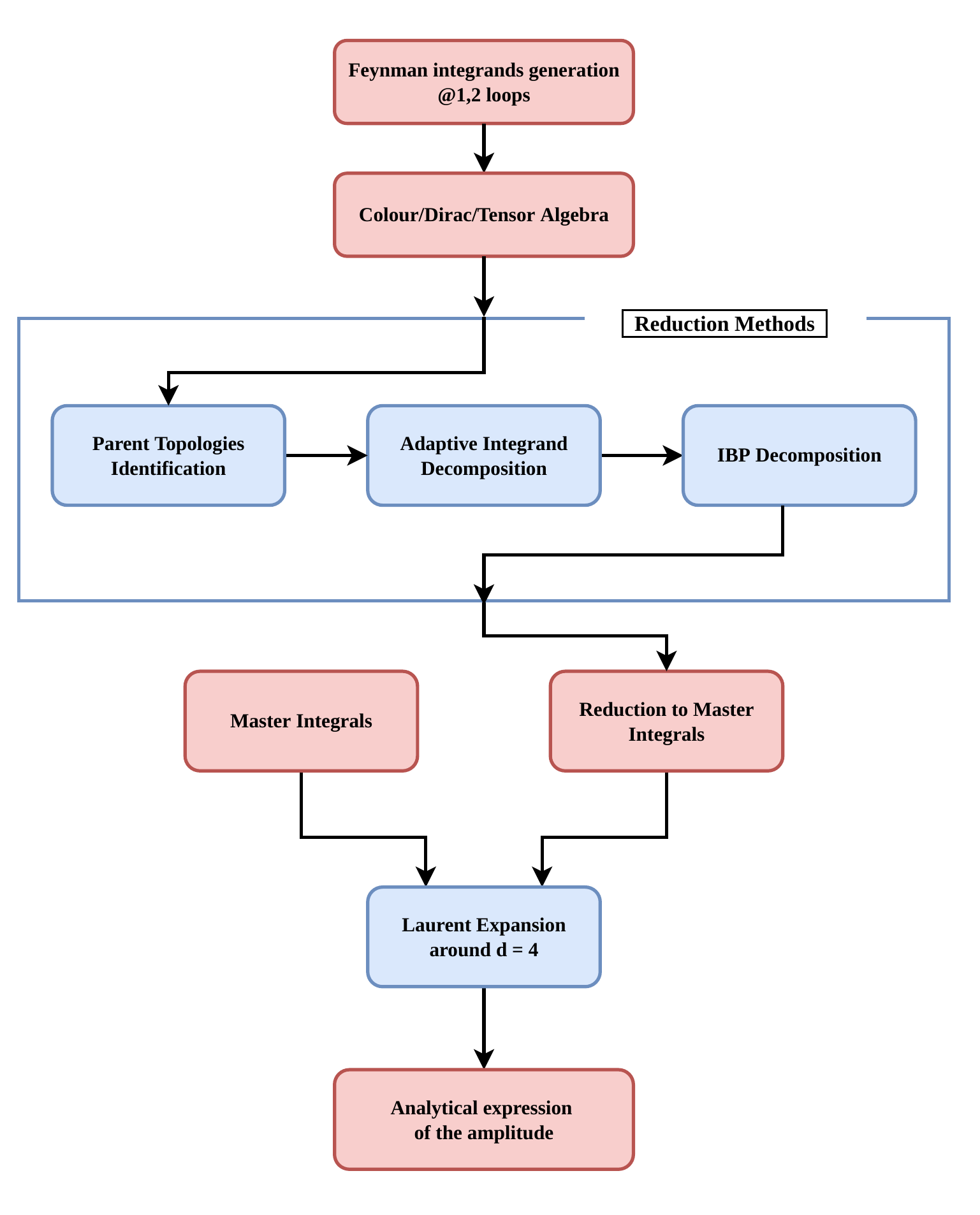}
    \caption{Flow-chart of the \Aida{} framework.} 
    \label{fig:flowchart}
\end{figure}

A flow chart of the complete computational algorithm implemented in the \Aida{} package is shown in 
Fig.~\ref{fig:flowchart}.

\subsection{IR structure}

The structure of IR singularities of the massless and massive gauge theory scattering amplitudes has been studied in~\cite{Catani:1998bh,Sterman:2002qn,Aybat:2006mz,Aybat:2006wq,Gardi:2009qi,Gardi:2009zv,Becher:2009cu,Becher:2019avh,Becher:2009qa,
Mitov:2006xs,
Becher:2009kw}.
The coefficients of the poles in $\epsilon$ appearing in the renormalised amplitudes ${\cal M}^{(1)}$ and ${\cal M}^{(2)}$ agree with the 
the universal IR structures of the QCD amplitudes, 
derived from the knowledge of the lower order terms, 
within SCET~\cite{Becher:2009qa, Becher:2009kw,:2012gk,Ferroglia:2009ep,Ferroglia:2009ii},
\begin{subequations}
\begin{eqnarray}
\label{eq:predpoles1L}
    \sum_{k=-2}^{-1} 
     {\cal M}^{(1)}_{k} \, \epsilon^k
   &=& 
   2 Z_{1}^\text{IR} \left.{\mathcal M}^{(0)}
  \right|_{\text{poles}} \, , \\
\label{eq:predpoles2L}
    \sum_{k=-4}^{-1} 
     {\cal M}^{(2)}_{k} \, \epsilon^k
     &=& 
     2 \Big[\!\!
     \left( Z_{2}^{\text{IR}} - \left(Z_{1}^{\text{IR}}\right)^2 \right) 
   {\mathcal M}^{(0)} 
   + \frac{1}{2} \, Z_{1}^{\text{IR}} \, {\mathcal M}^{(1)}
   \left. \Big] \right|_{\text{poles}} \, ,
    \qquad \quad
\end{eqnarray} 
\label{eq:predpoles}
\end{subequations}

\noindent
where $Z_i^{{\rm IR}}$ $(i=1,2)$ are the coefficients of the IR renormalisation factor $\boldsymbol{Z_{\text{IR}}}$ encoding the IR divergence. 
For the process under consideration, involving the production of a massive quark pair, $\boldsymbol{Z_{\text{IR}}}$ reads as~\cite{Ferroglia:2009ii},
\begin{eqnarray}
\boldsymbol{Z_{\text{IR}}} &=&
1 
+ \left(\frac{\alpha_s}{\pi} \right) Z_1^{{\rm IR}}
+ \left(\frac{\alpha_s}{\pi} \right)^2 Z_2^{{\rm IR}}
+ \mathcal{O}(\alpha_s^3) \ ,
\label{eq:IRRenomalizationFactor}
\end{eqnarray}
with,
\begin{eqnarray}
Z_1^{{\rm IR}} &=& 
\frac{\Gamma_0^\prime}{16\epsilon^2} + \frac{\boldsymbol\Gamma_0}{8\epsilon}  \,, \\
Z_2^{{\rm IR}} &=& 
\frac{(\Gamma_0^\prime)^2}{512\epsilon^4} + \frac{\Gamma_0^\prime}{128\epsilon^3}
\bigg( \boldsymbol\Gamma_0 - \frac{3}{2} \beta_0 \bigg) 
+ \frac{\boldsymbol\Gamma_0}{128\epsilon^2} 
\big( \boldsymbol\Gamma_0 - 2\beta_0 \big)
+ \frac{\Gamma_1^\prime}{256\epsilon^2}
+ \frac{\boldsymbol\Gamma_1}{64\epsilon} 
\nonumber \\
& &- \frac{2\, T_F}{3} \sum_{i=1}^{n_h} 
\bigg[ 
         \frac{\Gamma_0^\prime}{16}
         \bigg( \frac{1}{2\epsilon^2}\ln\frac{\mu^2}{m_i^2} + \frac{1}{4\epsilon}
         \bigg[\ln^2\frac{\mu^2}{m_i^2} +\frac{\pi^2}{6} \bigg] \bigg)
       + \frac{\boldsymbol\Gamma_0}{16\epsilon}
       \ln\frac{\mu^2}{m_i^2} 
\bigg]  
\,,
\end{eqnarray}
where  
$\Gamma_i' = \partial {\bf \Gamma_i}/\partial \ln \mu$, 
and $\boldsymbol \Gamma_i$ and $\beta_i$ 
are the coefficients of the perturbative expansion of the anomalous dimensions and of the QCD beta-function, respectively (expressed in the powers of renormalised coupling constant $\alpha_s$). The anomalous dimension matrix $\boldsymbol \Gamma$ for the $q\bar{q} \rightarrow t\bar{t}$ has been reported in~\cite{Ferroglia:2009ii}.

\subsection{Finite terms}  

\begin{figure}[t]
\centering
\includegraphics[scale=0.65]{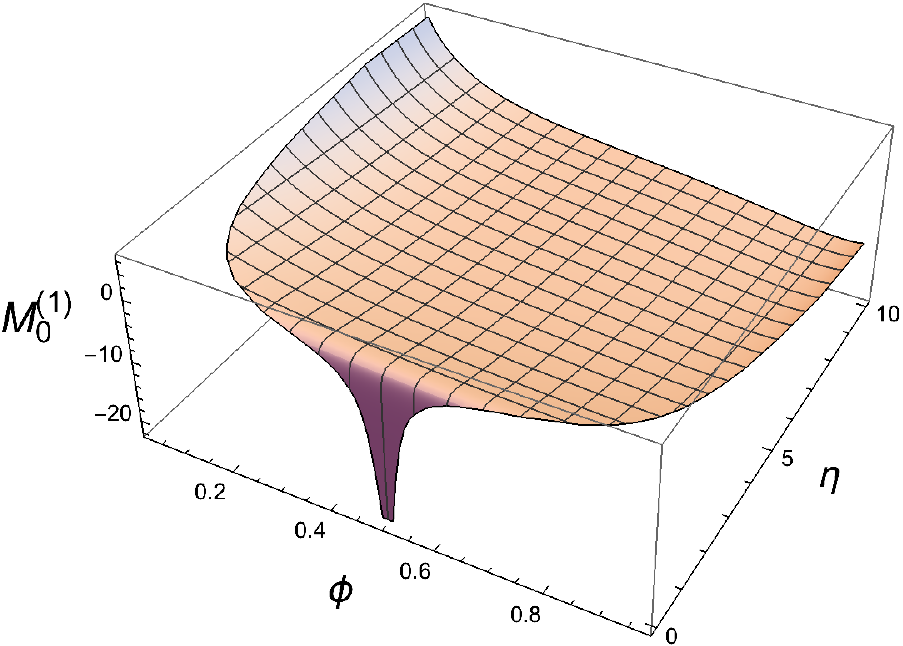}\qquad
\includegraphics[scale=0.65]{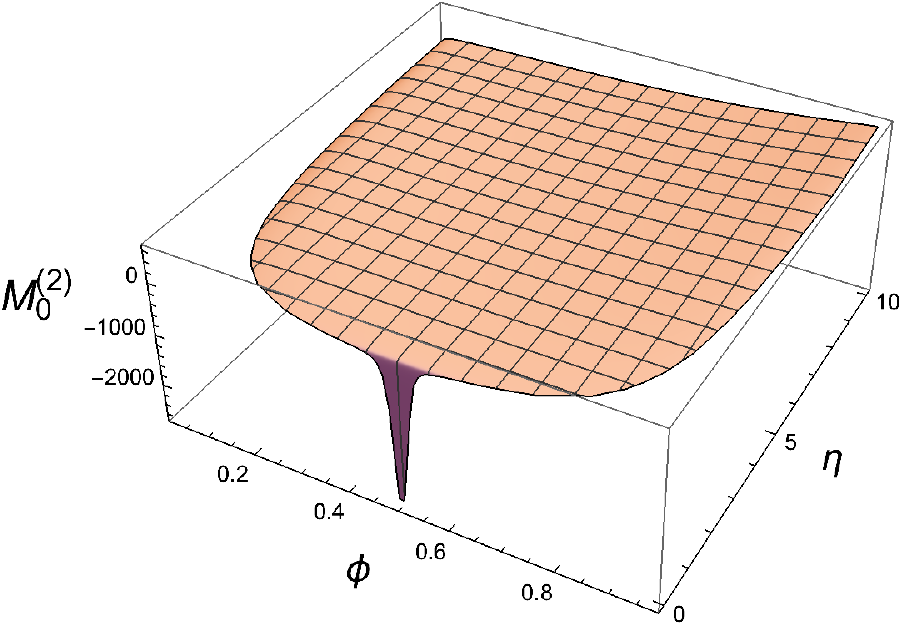}
\caption{Three-dimensional plots of the finite terms
${\cal M}_0^{(n)}$, $n=1,2$,
of the renormalised one- and two-loop amplitudes, in 
Eqs.~\eqref{eq:deco1L}, \eqref{eq:deco2L}, 
where $N_c=3, n_l=5$, and $n_h=1$.
}
\label{fig:3d_plots_total1L2L}
\end{figure}

\begin{figure}[b]
\centering
\includegraphics[scale=0.67]{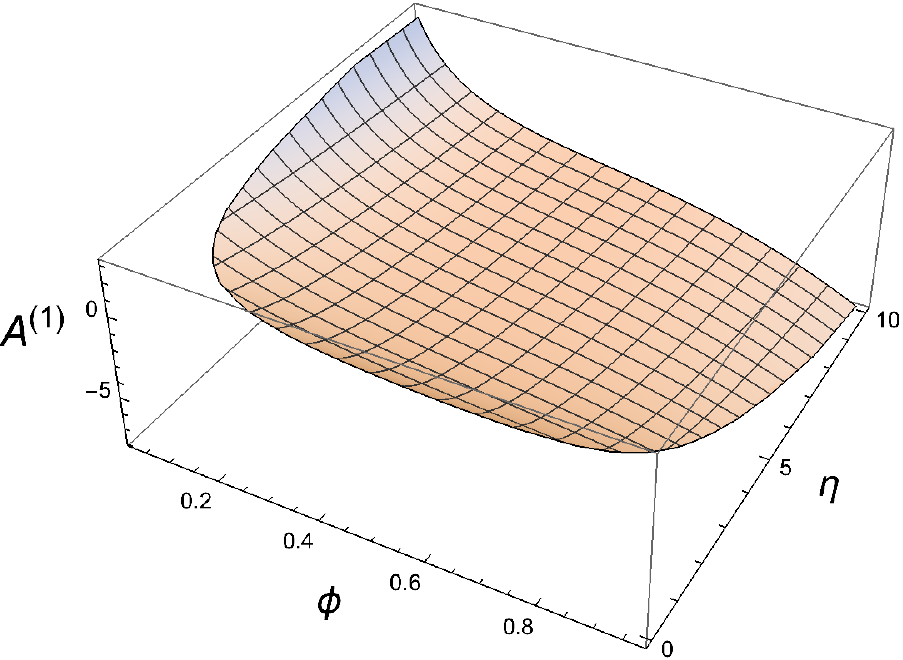}\qquad
\includegraphics[scale=0.67]{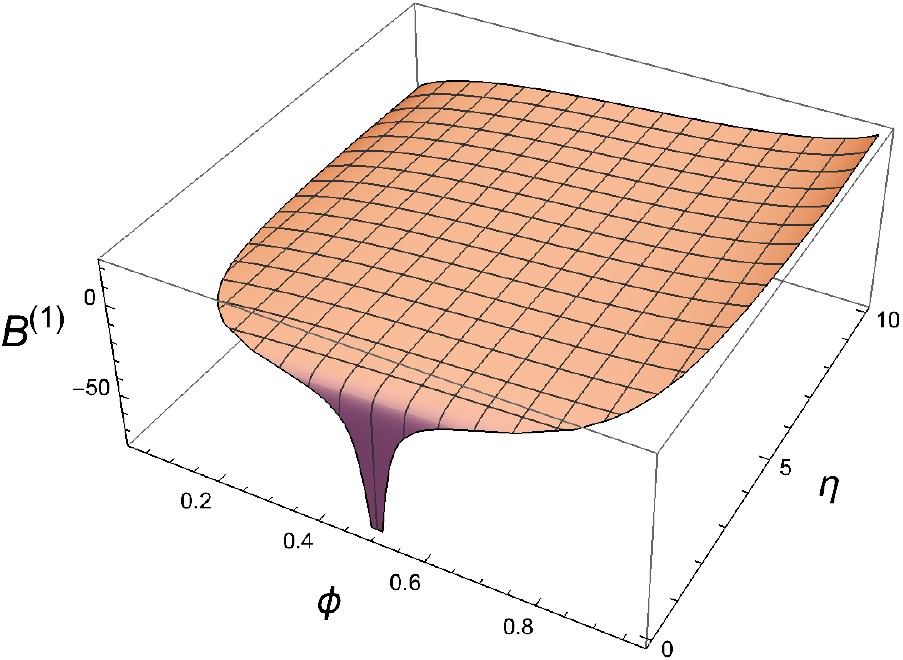}
\includegraphics[scale=0.67]{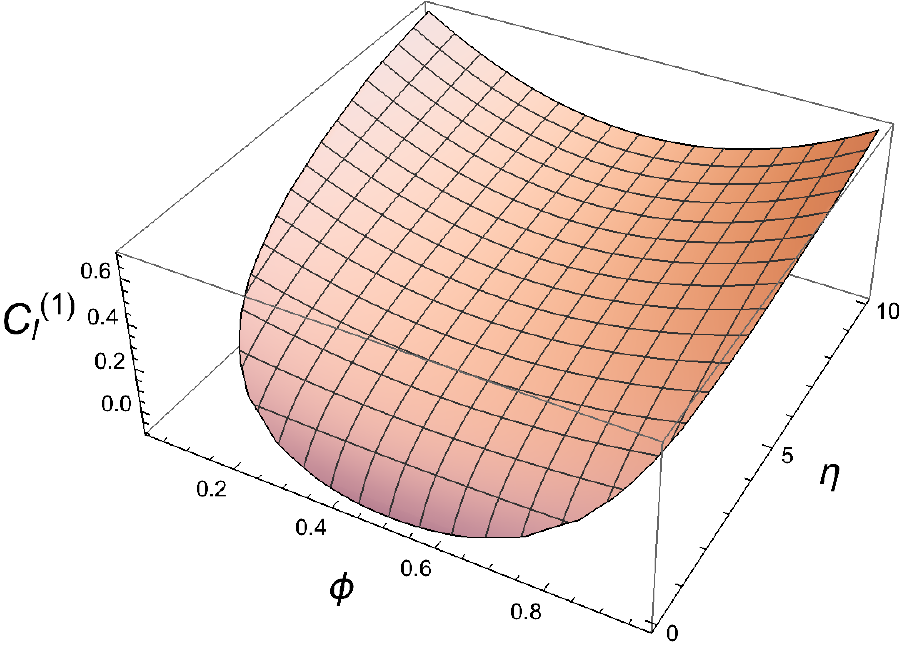}\qquad
\includegraphics[scale=0.67]{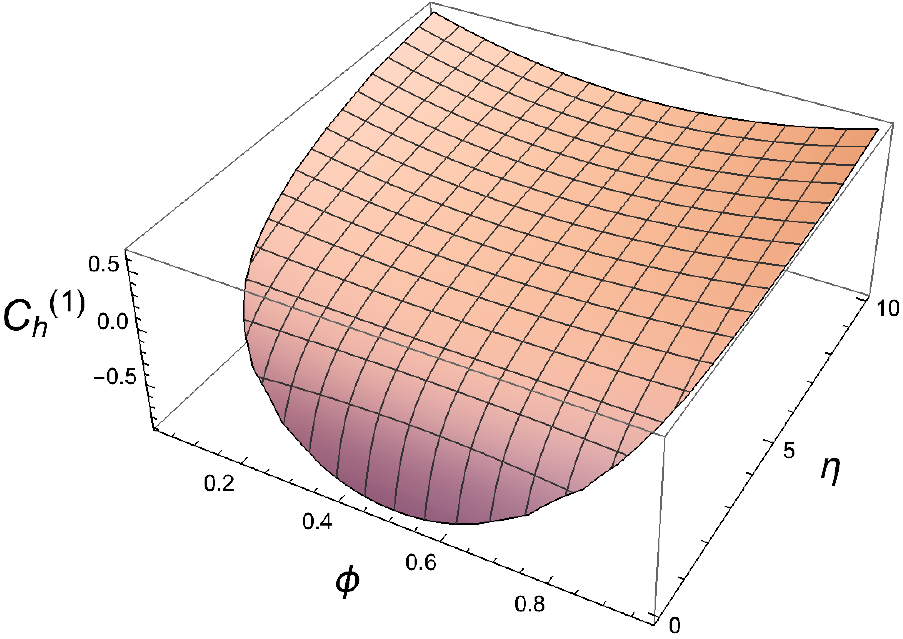}
\caption{Three-dimensional plots of the coefficients (finite part)
appearing in the decomposition of the renormalised one-loop amplitude in Eq.~\eqref{eq:deco1L}.
}
\label{fig:3d_plots_1L}
\end{figure}

\begin{figure}[h]
\centering
\includegraphics[scale=0.67]{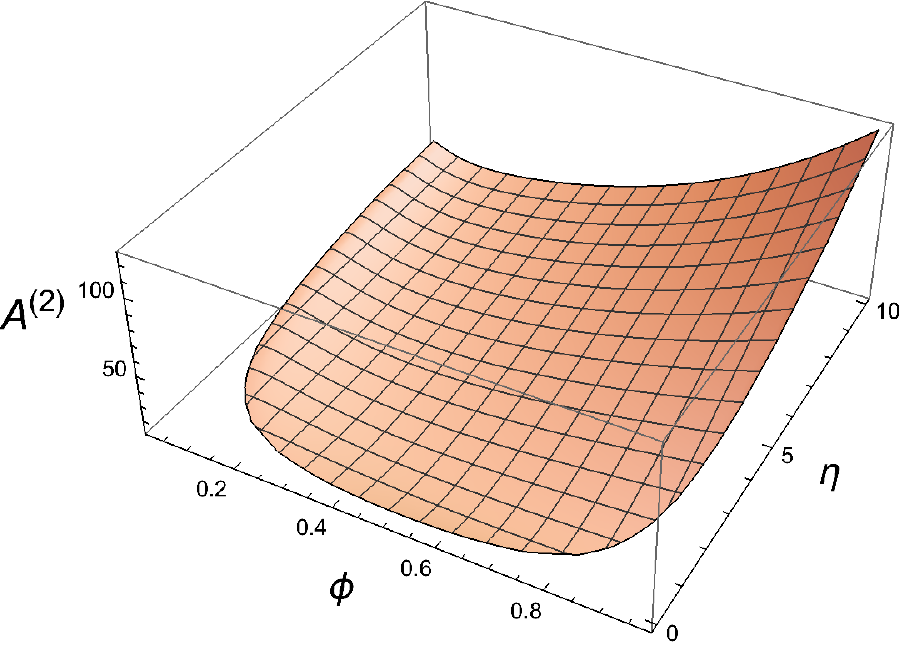}\qquad
\includegraphics[scale=0.67]{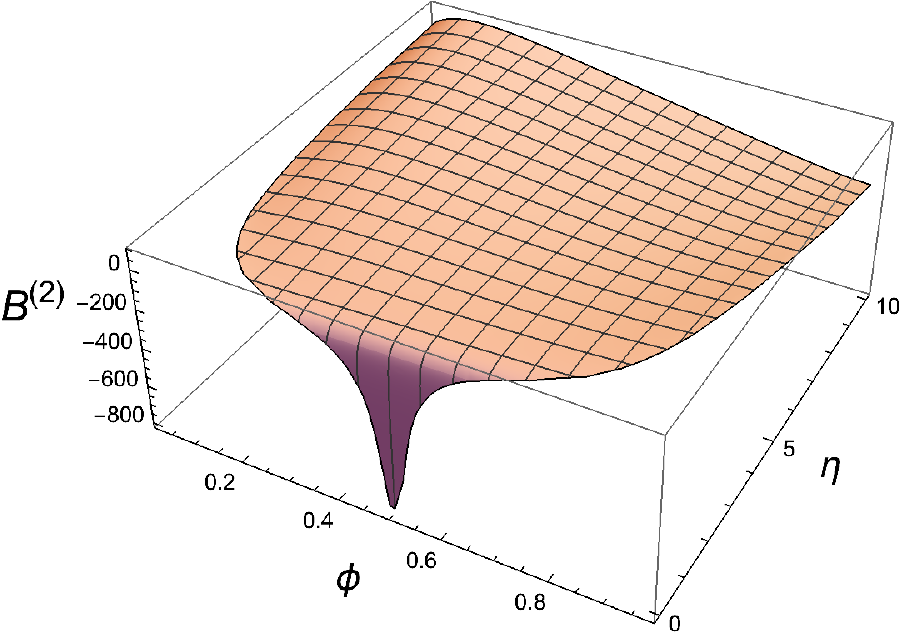}
\includegraphics[scale=0.67]{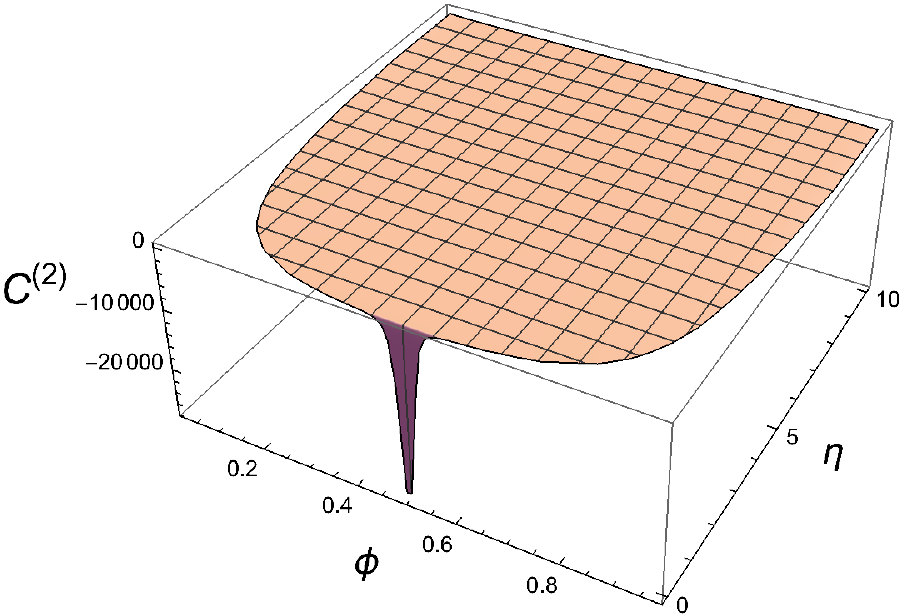}\qquad
\includegraphics[scale=0.67]{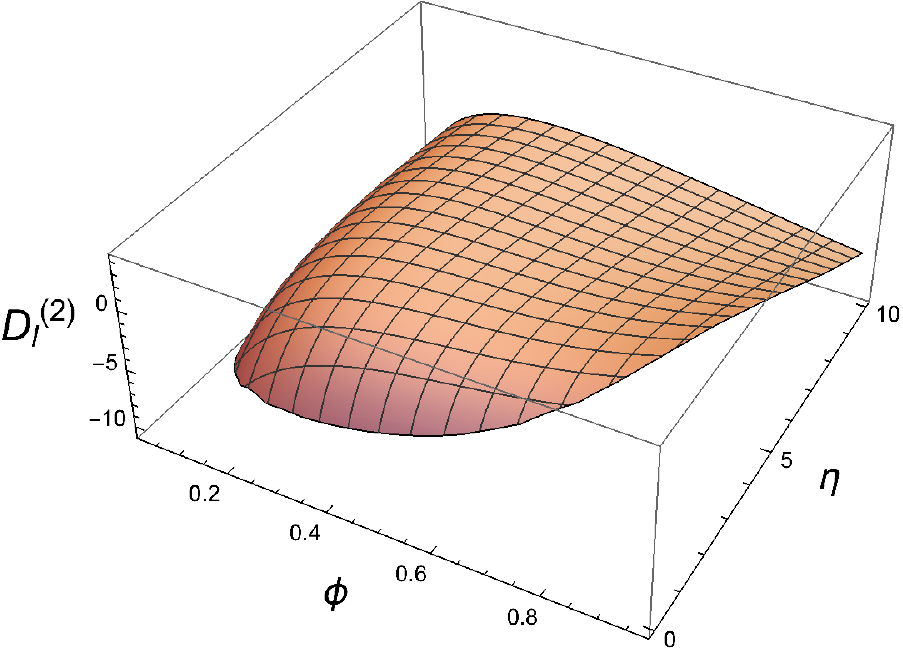}
\includegraphics[scale=0.67]{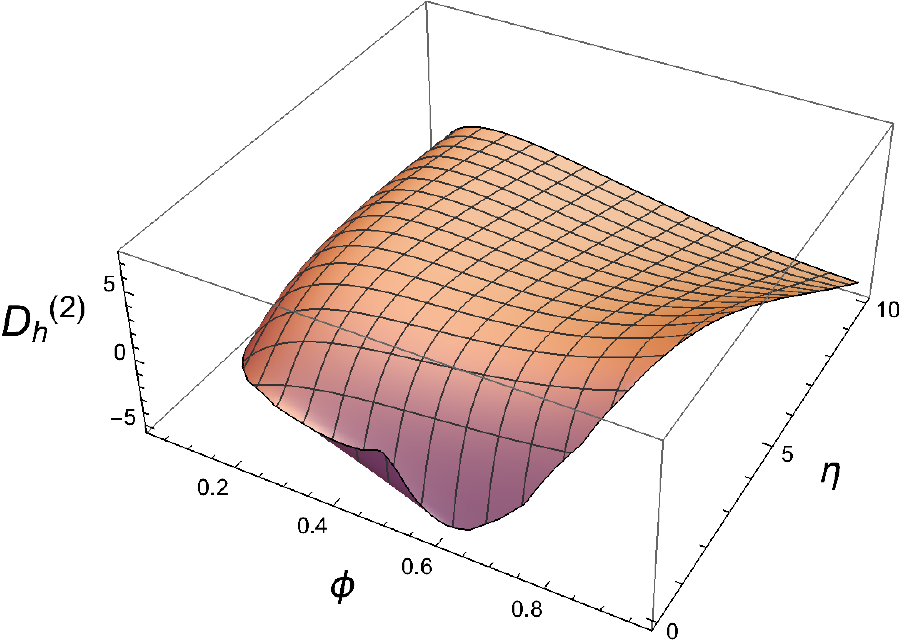}\qquad
\includegraphics[scale=0.67]{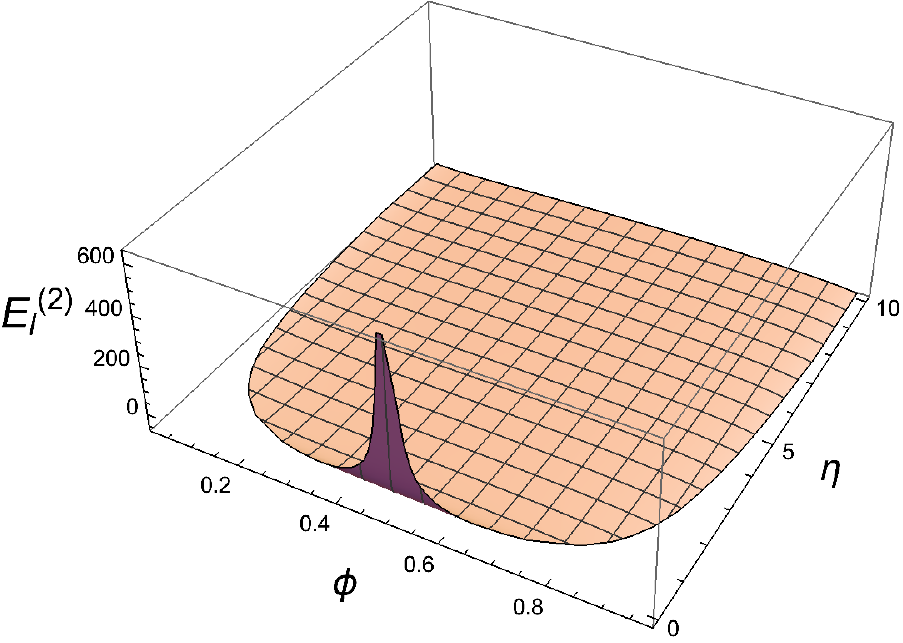}
\includegraphics[scale=0.67]{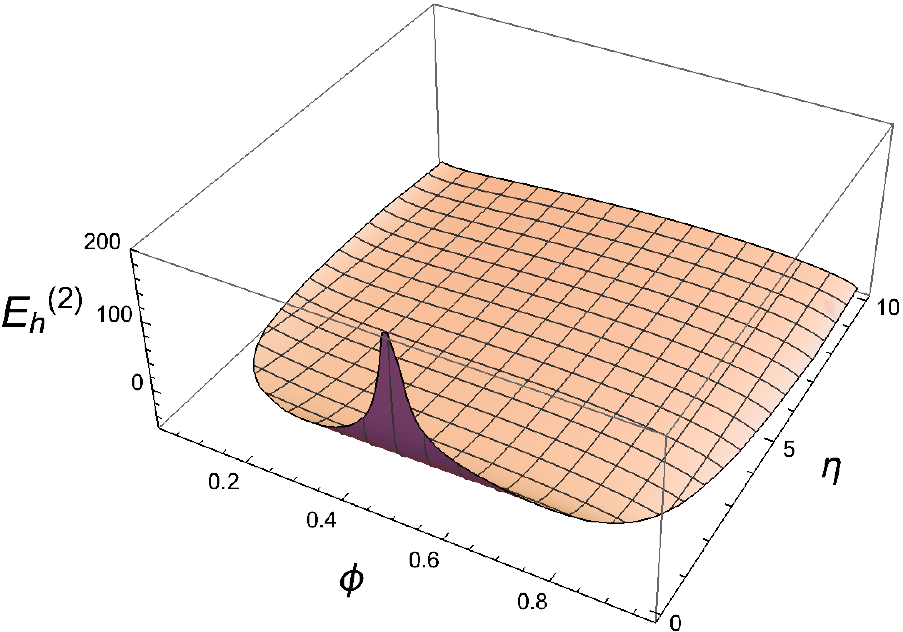}\qquad
\includegraphics[scale=0.67]{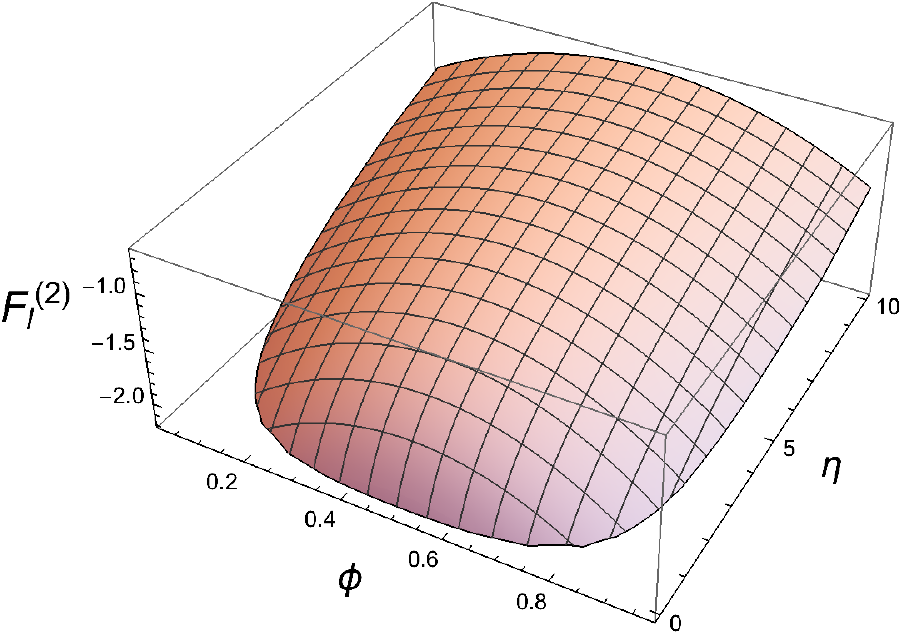}
\includegraphics[scale=0.67]{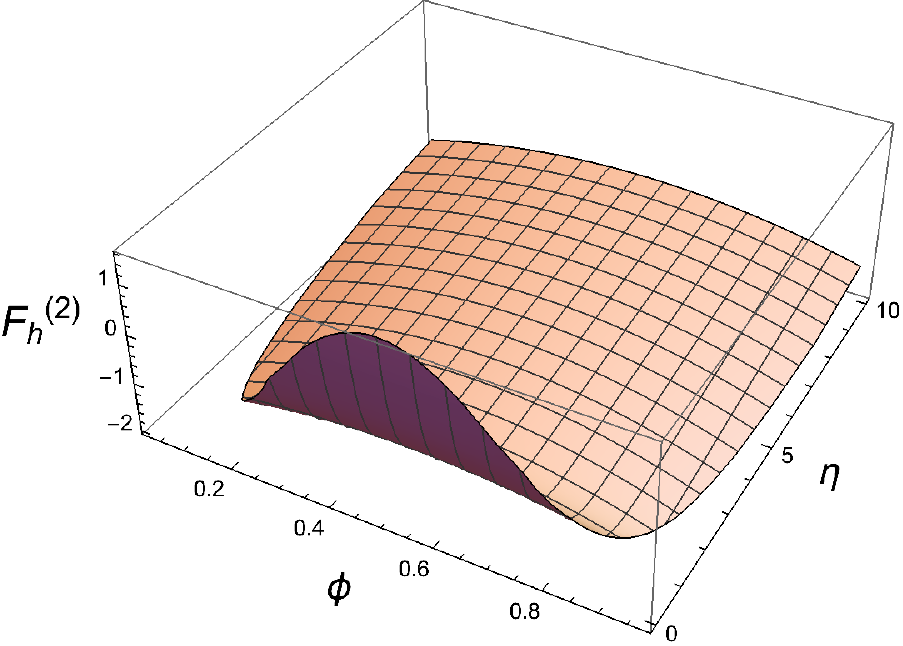}\qquad
\includegraphics[scale=0.67]{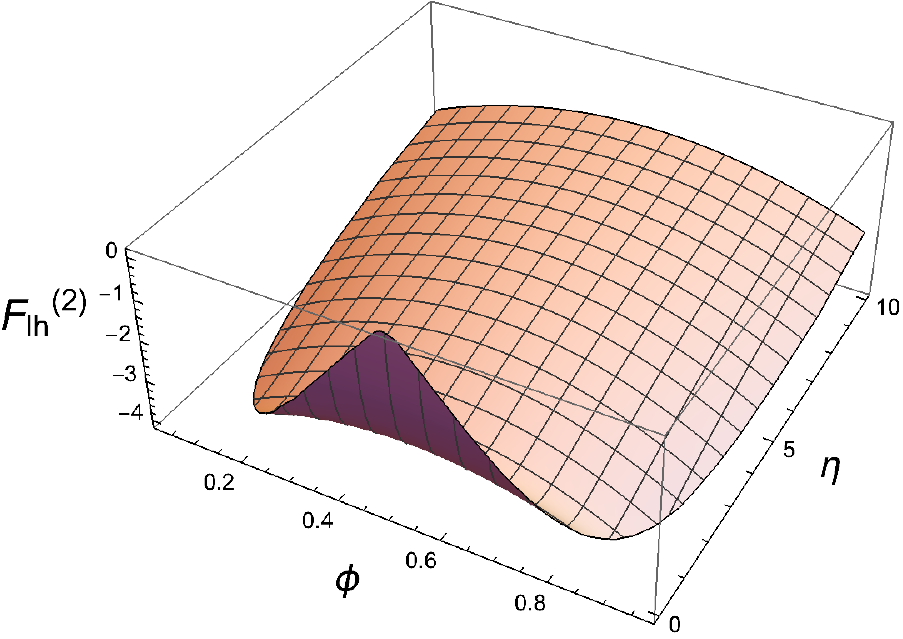}
\caption{Three-dimensional plots of the coefficients (finite part)
appearing in the decomposition of the renormalised two-loop amplitude in Eq.~\eqref{eq:deco2L}.
}
\label{fig:3d_plots_2L}
\end{figure}

In Fig.~\ref{fig:3d_plots_total1L2L}, we plot the finite part of one- and two-loop renormalised amplitudes ${\cal M}_0^{(i)}$, $i=1,2$ in the physical region,
as function of the auxiliary kinematic variables $\eta$ and $\phi$, defined in Eq.~\eqref{eq:physreg},
by setting $n_l = 5$, $n_h=1$, and $N_c=3$.
The contributions of the individual colour factors at one and two loops are shown in Figs. \ref{fig:3d_plots_1L} and~\ref{fig:3d_plots_2L}.

The plots are obtained by evaluating the analytic formulas at one and two loops 
with high precision on $10,\!500$ evenly spaced grid points.
 The numerical evaluation of the GPLs is carried out by {\sc HandyG}~\cite{Naterop:2019xaf} (away from threshold) and {\sc Ginac}~\cite{Vollinga:2004sn} (close to threshold) through their interface to {\sc PolyLogTools}~\cite{Duhr:2019tlz}.

The {\it finite term} of the analytic expression of the two-loop contribution ${\cal M}^{(2)}_0$, which constitutes the main result of this communication, is found in agreement with the numerical results available in~\cite{Czakon:2008zk}.
In particular, the numerical values of the grid attached to the arXiv submission of the latter reference agree with the (higher accuracy) values obtained from the numerical evaluation of our analytic expressions, in the same phase-space points.
For completeness, the values of 
${\cal M}_0^{(2)}$, numerically evaluated at 1600 phase-space points,
are given in the ancillary file 
\verb"qqQQGrid.m", 
attached to this publication.
Our grid is given in the format $\{ \phi, \eta, {\cal M}_0^{(2)} \}$, for the scale choice 
$\mu^2 = M^2$,
in the same phase-space points chosen in~\cite{Czakon:2008zk}.

In Table~\ref{tab:valuesofloopamp}, we showcase the numerical values of the analytic expressions of the individual colour factors, at one- and two-loop.
The analytic expressions are evaluated  with \Ginac{} 
at the kinematic point 
$s/M^2= 5$, $t/M^2=-5/4$, $\mu^2=M^2$
(following the prescription given in eq.(\ref{eq:deltaim}), the imaginary term is chosen to have $\delta=10^{-25}$),
which corresponds to 
the same kinematic point as in the Table 3 of~\cite{Czakon:2008zk} (see also
Table 1 of~\cite{Baernreuther:2013caa}), 
and our results are in agreement up-to the digits reported in the latter.

Moreover, the analytic expressions of the finite part, as well as of the poles, of $A^{(2)}$, $D_l^{(2)}$, $D_h^{(2)}$, $E_l^{(2)}$, $E_h^{(2)}$, $F_l^{(2)}, F_{lh}^{(2)}, F_{h}^{(2)}$
agree with earlier results published in~\cite{Bonciani:2008az,Bonciani:2009nb}.

\begin{landscape}
\begin{table}[b]
\begin{center}    
\begin{tabular}{|r|r|r|r|r|r|r|} 
\hline
    {} & {$\epsilon^{-4}$} 
       & {$\epsilon^{-3}$} 
       & {$\epsilon^{-2}$} 
       & {$\epsilon^{-1}$} 
       & {$\epsilon^{0}$} & {$\epsilon^{1}$}  \tabularnewline
\hline
    $A^{(0)}$ & - & - & - &  - &  $\frac{181}{100}$ & -2 \tabularnewline
\hline
\hline
$A^{(1)}$ & - & - & $-\frac{181}{400}$ &  0.1026418456757775  &  1.356145770566065  &  2.230403451742140 \tabularnewline
$B^{(1)}$ & - & - & $\frac{181}{400}$  &  -0.3180868339485723  &  -5.763132746701004  &  2.913169881363488 \tabularnewline
$C^{(1)}_{l}$ & - & - & 0  &  0  &  -0.01726400752682416  &  1.235821434465827 \tabularnewline
$C^{(1)}_{h}$ & - & - & 0  &  0  &  -0.5623350683773134  &  0.6373589172648111 \tabularnewline
\hline
\hline
    $A^{(2)}$ 
    & $\frac{181}{800}$  
    &  \underline{1.391733154}324222  
    &  \underline{-2.298174307}221209  
    &  \underline{-4.14575244}8999165  
    &  \underline{17.3713659}8564062 & - \tabularnewline
    $B^{(2)}$ 
    & $-\frac{181}{400}$  
    &  \underline{-1.323646320}375650  
    &  \underline{8.507455541}210568  
    &  \underline{6.035611156}200398  
    &  \underline{-35.12861106}350758 & - \tabularnewline
    $C^{(2)}$ 
    & $\frac{181}{800}$  
    &  \underline{-0.0680868339}4857230  
    &  \underline{-18.00716652}035224  
    &  \underline{6.302454931}016090  
    &  \underline{3.52404491}2826756 & - \tabularnewline
    $D_{l}^{(2)}$ 
    & 0  
    & - $\frac{181}{800}$ 
    &  \underline{0.260505733}8631945  
    &  \underline{-0.7250180282}219092  
    &  \underline{-1.93541724}6635768 & - \tabularnewline
    $D_{h}^{(2)}$ 
    &  0  
    &  0  
    &  \underline{0.562335068}3773134  
    &  \underline{0.1045606449}242690  
    &  \underline{-1.70474799}7587188 & - \tabularnewline
    $E_{l}^{(2)}$ 
    & 0  
    & $\frac{181}{800}$ 
    &  \underline{-0.3323207}299541260  
    &  \underline{7.904121951}420471  
    &  \underline{2.84869783}6597635 
    & - \tabularnewline
    $E_{h}^{(2)}$ & 0  &  0  
    &  \underline{-0.562335068}3773134  
    &  \underline{4.528240788}258799  
    &  \underline{12.73232424}278180 & - \tabularnewline
    $F_{l}^{(2)}$ 
    &  0  
    &  0  
    &  0  
    &  0  
    &  \underline{-1.984228442}234312 & - \tabularnewline
    $F_{lh}^{(2)}$ 
    &  0  
    &  0  
    &  0  
    &  0  
    &  \underline{-2.442562819}239786 & - \tabularnewline
    $F_{h}^{(2)}$ 
    &  0  
    &  0  
    &  0  
    &  0  
    &  \underline{-0.07924540546}146283 & - \tabularnewline
\hline\hline
\end{tabular}
\end{center}
    \caption{Numerical values of the 
    LO squared amplitude, in Eq.~\eqref{eq:bareborn}, and of the 
    coefficients appearing in the decomposition of the renormalised one- and two-loop amplitudes in Eqs.~\eqref{eq:deco1L}
    and~\eqref{eq:deco2L},
    evaluated at the phase space point $s/M^2= 5$, $t/M^2=-5/4$, $\mu^2=M^2$. 
    The underlined digits show the agreement with the results reported in Table 1 of~\cite{Baernreuther:2013caa}.}
    \label{tab:valuesofloopamp}
\end{table}
\end{landscape}

\section{Conclusion}
\label{sec:conclusion}
We completed the analytic evaluation of the scattering amplitude for the process $q {\bar q} \to Q {\bar Q}$ at two loops in QCD, for a massless $(q)$ and a massive $(Q)$ quark type. 
The contribution of the leading colour diagrams and of those containing fermion loops, 
whose analytical results were already available in the literature, 
were independently evaluated and cross checked, and combined with the novel contributions of the sub-leading colour terms, which were evaluated in this work, for the first time. 

The un-renormalised interference terms of the one- and two-loop bare amplitudes with the leading-order one were computed in the framework of CDR. 
The renormalisation of the ultraviolet divergences was carried out by employing the on-shell scheme for the quarks, 
and the $\overline{\text{MS}}$-scheme for the strong coupling constants. 

The analytic results of the one- and two-loop renormalised contributions, obtained as Laurent series around $d=4$ dimensions, respectively, up-to the first order term, and up-to the finite term,
were expressed in terms of GPLs and transcendental constants of up-to weight four. 
The singularity structure of the renormalised results was 
found to be in compliance with the predicted universal infrared behaviour of QCD amplitudes~\cite{Becher:2009kw,Ferroglia:2009ep,Ferroglia:2009ii}.
Numerical and partial analytical
results of the scattering amplitude already available in the literature~\cite{Czakon:2008zk,Bonciani:2008az,Bonciani:2009nb,Baernreuther:2013caa} agree with the novel analytic expression. 

The analytic results of the two-loop scattering amplitude for the
top-quark pair production from
the light-quark annihilation channel are an essential ingredient to be combined with the ones of the gluon fusion channel, whose analytic knowledge is partially available~\cite{Bonciani:2010mn,vonManteuffel:2013uoa,Bonciani:2013ywa,Adams:2018bsn,Adams:2018kez,Badger:2021owl}, to obtain -- hopefully, in a not-so-far future -- the full analytic expressions of the scattering amplitudes for the production of a heavy quark-antiquark pair in hadron collisions, at two loops in QCD~\cite{Czakon:2008zk,Baernreuther:2013caa}.

The results presented for the process $q {\bar q} \to Q {\bar Q}$ in QCD can be considered as an extension to the non-Abelian case of the ones recently obtained for the process $e^+ e^- \to \mu^+ \mu^-$ in QED~\cite{Bonciani:2021okt}. 
The automatic framework which was developed for these calculations is flexible and applicable to other scattering reactions. 
The computational efforts and the intermediate results for the non-Abelian case, such as diagram generation, integral and integrand decompositions, and evaluation of master integrals, are ingredients that are now available for the study of the elastic scattering processes of one massless and one massive particle/body, which is related by crossing symmetry to the one presented here. 

The competences acquired during this work, as well as the building blocks of the calculations, 
are not limited to applications within Particle Physics, and could be applied, for instance, to investigate processes in General Relativity, like the bending of light caused by a massive astrophysical body, see for instance~\cite{Bjerrum-Bohr:2016hpa,Bjerrum-Bohr:2017dxw}, where 
the massless quark is replaced by a photon, 
the massive quark is replaced by the world-line of a black-hole,
and gluons are replaced by gravitons.

Let us finally remark that, more generally, the presented results constitute a crucial reference for the study of the scattering of particles/bodies with non-vanishing masses, for interactions mediated by self-interacting massless quanta, in the limiting case when one of the body can be treated as massless. 
Therefore, they can offer additional insights for investigating similarities and differences between fundamental interactions occurring in different physical scenarios.


\acknowledgments 
We wish to thank A. Primo for collaboration at early stage of this project, in particular, during the development of \Aida{} and for discussions on the diagrams shown in Fig.~\ref{fig:zeronp}.
W.J.T. would like to thank J. Mazzitelli for suggesting numerical checks on the analytic expressions
presented in this manuscript. 
We wish to acknowledge 
R. Bonciani,
A. Broggio,
M. Czakon,
S. Di Vita,
A. Ferroglia,
F. Gasparotto,
T. Gehrmann, 
M. Grazzini,
A. Primo, 
U. Schubert,
A. Signer,
and F. Tramontano,
for stimulating discussions at various stages, and comments on the manuscript.
The work of M.K.M is supported by Fellini - Fellowship for Innovation at INFN funded by the European Union's Horizon 2020 research and innovation programme under the Marie Sk{\l}odowska-Curie grant agreement No 754496. 
J.R. acknowledges support from INFN. 
This project received funding from the European Research Council (ERC) under the European Union's Horizon 2020 research and innovation programme (grant agreement No 725110), {\it Novel structures in scattering amplitudes}. 

\appendix
\section{Colour Stripped Form Factors}
\label{app:colordeco}

The Feynman diagrammatic approach has been adopted throughout the calculation, and  
in this Appendix, we provide details on the contribution of the one- and two-loop Feynman diagrams to the 
form factors present in decompositions~\eqref{eq:deco1L} and~\eqref{eq:deco2L}, respectively. 

In decomposition~\eqref{eq:deco1L}, for the one-loop contribution,
we need to deal with 10 non-vanishing Feynman diagrams (see Fig.~\ref{fig:oneloopdiagsall}).
Two of them contain vacuum polarisation insertions
(with a closed heavy- and light-quark loop) 
contributing to form factors $C_l^{(1)}$ and $C_h^{(1)}$.
The remaining 8 diagrams may contribute to either 
$A^{(1)}$ (5 diagrams) or $B^{(1)}$ (4 diagrams).
In particular, 
$A^{(1)}$ gets contribution from purely planar 
diagrams with and without self-gluon interactions. 
$B^{(1)}, C_l^{(1)}$, and $C_h^{(1)}$ get contribution only from diagrams without self-gluon interactions.

Therefore, some of the form factors appearing in the decomposition of the considered amplitude, for 
a non-Abelian theory, can be written as linear combination of colour-stripped diagrams 
that would contribute to the scattering amplitude of an Abelian theory (like $e^{+} e^{-} \to \mu^{+} \mu^{-}$ in QED).
We list here, the decomposition of the form factors in terms of colour-stripped (Abelian-like) diagrams:
\begin{align}
B^{(1)} &= - {\rm d}^{(1)}_1 - {\rm d}^{(1)}_3 - 2{\rm d}^{(1)}_5 - 2{\rm d}^{(1)}_6 \ ,
\notag\\
C_l^{(1)} &= {\rm d}^{(1)}_7 \ ,
\notag\\
C_h^{(1)} &= {\rm d}^{(1)}_8 \ ,
\end{align}
where ${\rm d}_k^{(1)}$ accounts for colour-stripped $k$-th Feynman diagram of Fig.~\ref{fig:oneloopdiagsall}. 

Similarly, the form-factors appearing in decomposition of the two-loop amplitude in~\eqref{eq:deco2L}, 
gets contribution from 184 non-vanishing Feynman diagrams.
In particular:
$A^{(2)}$ gets contributions from 49 diagrams, which similarly to $A^{(1)}$, are only planar;
$B^{(2)}$ gets contributions from 62 (planar and non-planar) diagrams;
$C^{(2)}$ gets contributions from 35 (planar and non-planar) diagrams;
$D^{(2)}_l$, from 19 diagrams; 
$D^{(2)}_h$, from 20 diagrams; 
$E^{(2)}_l$,  from 15 diagrams; 
$E^{(2)}_h$, from 15 diagrams; 
$F^{(2)}_h$, from 1 diagram; 
$F^{(2)}_{lh}$, from 2 diagrams;
and  $F^{(2)}_l$, from 1 diagrams.

In the same way, as in the one-loop decomposition, we notice that 
form factors $A^{(2)}, B^{(2)}$, $D^{(2)}_l$ and $D^{(2)}_h$ get contributions from 
Feynman diagrams with and without self-gluon interactions,
whereas, $C^{(2)}, E_l^{(2)}, E_h^{(2)}, F^{(2)}_l, F^{(2)}_{lh}$, and $F^{(2)}_h$ 
contain only diagrams without self-gluon interaction.
Thus, the latter form factors can be decomposed in colour-stripped (Abelian-like) diagrams as:
\begin{equation}
    \begin{aligned}
    C^{(2)} \, =
        & \, {\rm d}^{(2)}_4+{\rm d}^{(2)}_{12}+{\rm d}^{(2)}_{17}+{\rm d}^{(2)}_{21}+{\rm d}^{(2)}_{29}+{\rm d}^{(2)}_{34}+{\rm d}^{(2)}_{38}+2 {\rm d}^{(2)}_{42}+3 {\rm d}^{(2)}_{44}+3 {\rm d}^{(2)}_{45}+2 {\rm d}^{(2)}_{46} \\
	    & +2 {\rm d}^{(2)}_{48}+2 {\rm d}^{(2)}_{49}+2 {\rm d}^{(2)}_{51}+2 {\rm d}^{(2)}_{53}+2 {\rm d}^{(2)}_{54}+2 {\rm d}^{(2)}_{56}+2 {\rm d}^{(2)}_{58}+2 {\rm d}^{(2)}_{60}+2 {\rm d}^{(2)}_{62} \\
	    & +3 {\rm d}^{(2)}_{64}+3 {\rm d}^{(2)}_{65}+2 {\rm d}^{(2)}_{66}+3 {\rm d}^{(2)}_{68}+3 {\rm d}^{(2)}_{69}+{\rm d}^{(2)}_{106}+{\rm d}^{(2)}_{107}+{\rm d}^{(2)}_{112}+{\rm d}^{(2)}_{130} \\
	    & +{\rm d}^{(2)}_{131}+{\rm d}^{(2)}_{136}+2 {\rm d}^{(2)}_{158}+2 {\rm d}^{(2)}_{163}+2 {\rm d}^{(2)}_{164}+2 {\rm d}^{(2)}_{165} \ , \\
    E^{(2)}_{\rm l}\, =
        & \, -2 {\rm d}^{(2)}_8-2 {\rm d}^{(2)}_{10}-2 {\rm d}^{(2)}_{25}-2 {\rm d}^{(2)}_{27}-{\rm d}^{(2)}_{78}-{\rm d}^{(2)}_{88}-{\rm d}^{(2)}_{90}-{\rm d}^{(2)}_{98} \\
	    & -{\rm d}^{(2)}_{117}-{\rm d}^{(2)}_{122}-{\rm d}^{(2)}_{141}-2 {\rm d}^{(2)}_{146}-2 {\rm d}^{(2)}_{150}-2 {\rm d}^{(2)}_{154}-2 {\rm d}^{(2)}_{159} \ , \\
    E^{(2)}_{\rm h}\, =
        & \, -2 {\rm d}^{(2)}_9-2 {\rm d}^{(2)}_{11}-2 {\rm d}^{(2)}_{26}-2 {\rm d}^{(2)}_{28}-{\rm d}^{(2)}_{79}-{\rm d}^{(2)}_{89}-{\rm d}^{(2)}_{91}-{\rm d}^{(2)}_{99}\\
    	& -{\rm d}^{(2)}_{118}-{\rm d}^{(2)}_{123} - {\rm d}^{(2)}_{142}-2 {\rm d}^{(2)}_{147}-2 {\rm d}^{(2)}_{151}-2 {\rm d}^{(2)}_{155}-2 {\rm d}^{(2)}_{160} \ , \\
    F^{(2)}_{\rm l}\, =
        & \, {\rm d}^{(2)}_{168} \ , \\
    F^{(2)}_{\rm lh}\, =
        & \, {\rm d}^{(2)}_{169}+{\rm d}^{(2)}_{170} \ , \\
    F^{(2)}_{\rm h}\, =
        & \, {\rm d}^{(2)}_{171} \ ,
    \end{aligned}
\end{equation}
with ${\rm d}_k^{(2)}$ stands for the colour-stripped $k$-th Feynman diagram of Figs.~\ref{fig:2Ldiagrams_part1}
and~\ref{fig:2Ldiagrams_part2}.  

\section{Renormalisation Constants}
\label{app:renconstants}

\noindent
In this Appendix, we provide the expressions of the UV renormalisation constants introduced in Sec. \ref{sec:UVRenormalization}, for convenience: \\

\noindent
$\bullet$ Light quark field:
\begin{eqnarray}
\delta Z_q^{(1)} &=& 0 \ ; \\
\delta Z_q^{(2)} &=& C_f T_f \, n_h
\Bigg(
  \frac{1}{16\ep}
  + \frac{1}{8} L_{\mu}{}
  - \frac{5}{96}
\Bigg) \ ;
\end{eqnarray}

\noindent
$\bullet$ Heavy quark field and mass:
\begin{eqnarray}
\delta Z_{Q}^{(1)} &=& C_f \Bigg(
- \frac{3}{4\ep}
- 1
- \frac{3}{4} L_{\mu}{}
+ \ep \Big(
- 2
- L_{\mu}{}
- \frac{3}{8}  L_{\mu}^2
- \frac{\pi^2}{16}
\Big)
+ \ep^2 \Big(
- 4 
- 2 L_{\mu}{}
- \frac{1}{2} L_{\mu}^2
\nonumber
\\
&-& \frac{1}{8} L_{\mu}^3
- \frac{\pi^2}{12}
- \frac{\pi^2}{16} L_{\mu}{}
+ \frac{1}{4} \zeta_3
\Big)
\Bigg) \ ;
 \\
\delta Z_{Q}^{(2)} &=& 
C_f T_f \, n_h \Bigg(
\frac{1}{16\ep}
+ \frac{1}{4\ep} L_{\mu}{}
+ \frac{947}{288}
+ \frac{11}{24} L_{\mu}{}
+ \frac{3}{8} L_{\mu}^2
- \frac{5\pi^2}{16}
\Bigg)
\nonumber \\
&+&
C_f T_f \, n_l \Bigg(
- \frac{1}{8\ep^2}
+ \frac{11}{48\ep}
+ \frac{113}{96}
+ \frac{19}{24} L_{\mu}{}
+ \frac{1}{8} L_{\mu}^2
+ \frac{\pi^2}{12}
\Bigg)
\nonumber \\
&+& C_f^2 \Bigg(
\frac{9}{32\ep^2}
+ \frac{51}{64\ep}
+ \frac{9}{16\ep} L_{\mu}{}
+ \frac{433}{128}
+ \frac{51}{32} L_{\mu}{}
+ \frac{9}{16} L_{\mu}^2
- \frac{49\pi^2}{64}
+ \pi^2 \ln(2) 
- \frac{3 \zeta_3}{2} 
\Bigg)
\nonumber \\
&+& C_f C_A \Bigg(
\frac{11}{32\ep^2}
- \frac{127}{192\ep}
- \frac{1705}{384}
- \frac{215}{96} L_{\mu}{}
- \frac{11}{32} L_{\mu}^2
+ \frac{5\pi^2}{16}
- \frac{\pi^2 \ln(2)}{2}  
+ 3 \zeta_3
\Bigg) \ ;
\\
\delta Z_{M}^{(1)} &=& C_f \Bigg(
- \frac{3}{4\ep}
- 1
- \frac{3}{4} L_{\mu}{}
+ \ep \Big(
- 2
- L_{\mu}{}
- \frac{3}{8}  L_{\mu}^2
- \frac{\pi^2}{16}
\Big)
+ \ep^2 \Big(
- 4 
- 2 L_{\mu}{}
- \frac{1}{2} L_{\mu}^2
\nonumber
\\
&-& \frac{1}{8} L_{\mu}^3
- \frac{\pi^2}{12}
- \frac{\pi^2}{16} L_{\mu}{}
+ \frac{1}{4} \zeta_3
\Big)
\Bigg) ; 
\end{eqnarray}

\noindent
$\bullet$ Coupling constant:
\begin{eqnarray}
\delta Z_{\alpha_s}^{(1)} &=&
\Bigg(
-\frac{11 }{12 \ep} C_A
+ {1 \over 3 \ep} T_f \left(n_l +n_h\right
)\Bigg) \ ;
\\
\delta Z_{\alpha_s}^{(2)} &=&
C_A^2 
\left(
\frac{121}{144 \ep^2}
-\frac{17}{48 \ep}
\right)
+C_A T_f (n_l+n_h) 
\left(
\frac{5}{24 \ep}
-\frac{11}{18\ep^2}
\right)
+ C_f T_f (n_l+n_h)  \frac{1}{8 \ep}
\nonumber \\
&+& T_f^2  \left(n_l+n_h\right)^2 \frac{1}{9 \ep^2} \ ;
\end{eqnarray}
where $L_\mu \equiv \ln(\mu^2/M^2)$.

\bibliographystyle{JHEP}
\bibliography{refs1}

\end{document}